\documentclass[final,smallextended]{svjour3}

\usepackage{amsmath}
\usepackage{amssymb}
\usepackage{graphicx}
\usepackage{pict2e}
\usepackage{color}
\usepackage{hyperref}
\usepackage[headings]{fullpage}

\newcommand{\C}[2]{{#1 \choose #2}}
\renewcommand{\O}{\operatorname{\mathcal{O}}}
\newcommand{\Li}{\mathrm{Li}}
\newcommand{\sgn}{\operatorname{sgn}}
\renewcommand{\Re}{\operatorname{Re}}
\renewcommand{\Im}{\operatorname{Im}}
\newcommand{\openone}{\mathbf{1}}
\newcommand{\half}{\tfrac{1}{2}}
\newcommand{\mint}{\approx\!\!\!\!\!\!\!\int}

\newcommand{\rme}{\mathrm{e}}
\newcommand{\rmi}{\mathrm{i}}
\newcommand{\rmd}{\mathrm{d}}

\definecolor{lightblue}{rgb}{0.7,0.7,1}
\definecolor{lightgray}{rgb}{0.8,0.8,0.8}

\begin{document}
\title{Asymptotics for the norm of Bethe eigenstates in the periodic totally asymmetric exclusion process}
\titlerunning{Norm of Bethe eigenstates for periodic TASEP: asymptotics}
\author{Sylvain Prolhac}
\institute{S. Prolhac \at Laboratoire de Physique Th\'eorique, IRSAMC, UPS, Universit\'e de Toulouse, France\\Laboratoire de Physique Th\'eorique, UMR 5152, Toulouse, CNRS, France\\\email{sylvain.prolhac@irsamc.ups-tlse.fr}}
\date{\today}

\maketitle
\begin{picture}(0,0)(10,-65)
\put(0,0){\color{white}\polygon*(0,0)(170,0)(170,20)(0,20)}
\end{picture}

\begin{abstract}
The normalization of Bethe eigenstates for the totally asymmetric simple exclusion process on a ring of $L$ sites is studied, in the large $L$ limit with finite density of particles, for all the eigenstates responsible for the relaxation to the stationary state on the KPZ time scale $T\sim L^{3/2}$. In this regime, the normalization is found to be essentially equal to the exponential of the action of a scalar free field. The large $L$ asymptotics is obtained using the Euler-Maclaurin formula for summations on segments, rectangles and triangles, with various singularities at the borders of the summation range.

\end{abstract}

\begin{section}{Introduction}
\label{section introduction}
Understanding the large scale evolution of macroscopic systems from their microscopic dynamics is one of the central aims of statistical physics out of equilibrium. Much progress has been happening toward this goal for systems in the one-dimensional KPZ universality class \cite{HHZ1995.1,SS2010.4,KK2010.1,C2011.1}, which describes the fluctuations in some specific regimes for the height of the interface in growth models, the current of particles in driven diffusive systems and the free energy for directed polymers in random media.

The totally asymmetric simple exclusion process (TASEP) \cite{D1998.1,GM2006.1} belongs to KPZ universality. On the infinite line, the current fluctuations in the long time limit \cite{J2000.1} are equal to the ones that have been obtained from other models, in particular polynuclear growth model \cite{PS2000.2}, directed polymer in random media \cite{BCP2014.1}, and from the Kardar-Parisi-Zhang equation \cite{KPZ1986.1} itself using the replica method \cite{CLDR2010.1,D2010.1}. On a finite system, the stationary large deviations of the current for periodic TASEP \cite{DL1998.1} agree with the ones from the replica method \cite{BD2000.1} and with the ones for open TASEP at the transition separating the maximal current phase with the high and low density phases \cite{GLMV2012.1}.

Much less is currently known about the crossover between fluctuations on the infinite line and in a finite system, see however \cite{LK2006.1,BPPP2006.1,GMGB2007.1,B2009.1,PBE2011.1,MSS2012.1,MSS2012.2}. The crossover takes place on the relaxation scale with times $T$ of order $L^{3/2}$ characteristic of KPZ universality in $1+1$ dimension. The aim of the present paper is to compute the large $L$ limit of the normalization of the Bethe eigenstates of TASEP that contribute to the relaxation regime. Our main result is that this limit depends on the eigenstate essentially through the free action of a field $\varphi$ built by summing over elementary excitations corresponding to the eigenstate. This result can be used to derive an exact formula for the current fluctuations in the relaxation regime \cite{P2014.2}.

The paper is organized as follows. In section \ref{section TASEP}, we briefly recall the master equation generating the time evolution of TASEP and its deformation which allows to count the current of particles. In section \ref{section Bethe ansatz}, we summarize some known facts about Bethe ansatz for periodic TASEP, and state our main result (\ref{asymptotics norm}) about the asymptotics of the norm of Bethe eigenstates. In section \ref{section Euler-Maclaurin}, we state the Euler-Maclaurin formula for summation on segments, triangles and rectangles with various singularities at the borders of the summation range. The Euler-Maclaurin formula is then used in section \ref{section asymptotics} to compute the asymptotics of the normalization of Bethe states. In appendix \ref{appendix zeta}, some properties of simple and double Hurwitz zeta functions are summarized.
\end{section}

\begin{section}{Periodic TASEP}
\label{section TASEP}
We consider TASEP with $N$ hard-core particles on a periodic lattice of $L$ sites. The continuous time dynamics consists of particle hopping from any site $i$ to the next site $i+1$ with rate $1$ if the destination site is empty.

Since TASEP is a Markov process, the time evolution of the probability $P_{T}(\mathcal{C})$ to observe the system at time $T$ in the configuration $\mathcal{C}$ is generated by a master equation. A deformation of the master equation can be considered \cite{DL1998.1} to count the total integrated current of particles $Q_{T}$, defined as the total number of hops of particles up to time $T$. Defining $F_{T}(\mathcal{C})=\sum_{Q=0}^{\infty}\rme^{\gamma Q}P_{T}(\mathcal{C},Q)$ where $P_{T}(\mathcal{C},Q)$ is the joint probability to have the system in configuration $\mathcal{C}$ with $Q_{T}=Q$, one has
\begin{equation}
\label{master eq F}
\frac{\partial}{\partial T}\,F_{T}(\mathcal{C})=\sum_{\mathcal{C}'\neq\mathcal{C}}\Big[\rme^{\gamma}w(\mathcal{C}\leftarrow\mathcal{C}')F_{T}(\mathcal{C}')-w(\mathcal{C}'\leftarrow\mathcal{C})F_{T}(\mathcal{C})\Big]\;.
\end{equation}
The hopping rate $w(\mathcal{C}'\leftarrow\mathcal{C})$ is equal to $1$ if the configuration $\mathcal{C}'$ can be obtained from $\mathcal{C}$ by moving one particle from a site $i$ to $i+1$, and is equal to $0$ otherwise. The deformed master equation (\ref{master eq F}) reduces to the usual master equation for the probabilities $P_{T}(\mathcal{C})$ when the fugacity $\gamma$ is equal to $0$. It can be encoded in a deformed Markov operator $M(\gamma)$ acting on the configuration space of dimension $\Omega=\C{L}{N}$ in the sector with $N$ particles. Gathering the $F_{T}(\mathcal{C})$ in a vector $|F_{T}\rangle$, one can write
\begin{equation}
\label{master eq |F>}
\partial_{T}|F_{T}\rangle=M(\gamma)F_{T}\;.
\end{equation}

The deformed master equation (\ref{master eq F}), (\ref{master eq |F>}) is known \cite{DL1998.1} to be integrable in the sense of quantum integrability, also called stochastic integrability \cite{S2012.1} in the context of an evolution generated by a non-Hermitian stochastic operator. At $\gamma=0$, the eigenvalue of the first excited state (gap) has been shown to scale as $L^{-3/2}$ using Bethe ansatz \cite{GS1992.1,GM2004.1,GM2005.1}. The whole spectrum has also been studied \cite{P2013.1}, and in particular the region with eigenvalues scaling as $L^{-3/2}$ \cite{P2014.1}. In this article, we study the normalization of the corresponding eigenstates, which is needed for the calculation of fluctuations of $Q_{T}$ on the relaxation scale $T\sim L^{3/2}$ \cite{P2014.2}. There
\begin{equation}
\label{GF[M]}
\langle\rme^{\gamma Q_{T}}\rangle=\frac{\langle\mathcal{C}|\rme^{TM(\gamma)}|\mathcal{C}_{0}\rangle}{\langle\mathcal{C}|\rme^{TM}|\mathcal{C}_{0}\rangle}\;
\end{equation}
is evaluated by inserting a decomposition of the identity operator in terms of left and right normalized eigenvectors. Throughout the paper, we consider the thermodynamic limit $L,N\to\infty$ with fixed density of particles
\begin{equation}
\label{rho}
\rho=\frac{N}{L}\;,
\end{equation}
and fixed rescaled fugacity
\begin{equation}
\label{s}
s=\sqrt{\rho(1-\rho)}\,\gamma\,L^{3/2}\;
\end{equation}
according to the relaxation scale $T\sim L^{3/2}$ in one-dimensional KPZ universality.
\end{section}

\begin{section}{Bethe ansatz}
\label{section Bethe ansatz}
In this section, we recall some known facts about Bethe ansatz for periodic TASEP, and state our main result about the asymptotics of the normalization of Bethe eigenstates.

\begin{subsection}{Eigenvalues and eigenvectors}
Bethe ansatz is one of the main tools that have been used to obtain exact results about dynamical properties of TASEP. It allows to diagonalize the $N$ particle sector of the generator of the evolution $M(\gamma)$ in terms of $N$ (complex) momenta $q_{j}$, $j=1,\ldots,N$. The eigenvectors are then written as sums over all $N!$ permutations assigning momenta to the particles. On the infinite line, the momenta are integrated over on some continuous curve in the complex plane \cite{TW2008.1}. For a finite system on the other hand, only a discrete set of $N$-tuples of momenta are allowed, as is usual for particles in a box. Writing $y_{j}=1-\rme^{\gamma}\rme^{\rmi q_{j}}$, one can show that for the system with periodic boundary conditions, the complex numbers $y_{j}$, $j=1,\ldots,N$ have to satisfy the \textit{Bethe equations}
\begin{equation}
\label{BE[y]}
\rme^{L\gamma}(1-y_{j})^{L}=(-1)^{N-1}\prod_{k=1}^{N}\frac{y_{j}}{y_{k}}\;.
\end{equation}
We use the shorthand $r$ to refer to the sets of $N$ \textit{Bethe roots} $y_{j}$ solving the Bethe equations. The eigenstates are indexed by $r$. The corresponding eigenvalue of $M(\gamma)$ is equal to
\begin{equation}
\label{E[y]}
E_{r}(\gamma)=\sum_{j=1}^{N}\frac{y_{j}}{1-y_{j}}\;.
\end{equation}
By translation invariance of the model, each eigenstate of $M(\gamma)$ is also eigenstate of the translation operator. The corresponding eigenvalue is
\begin{equation}
\rme^{2\rmi\pi p_{r}/L}=\rme^{N\gamma}\prod_{j=1}^{N}(1-y_{j})\;,
\end{equation}
with total momentum $p_{r}\in\mathbb{Z}$.

The coefficients of the right and left (unnormalized) eigenvectors for a configuration with particles at positions $1\leq x_{1}<\ldots<x_{N}\leq L$ are given by the determinants
\begin{eqnarray}
\label{psiR[y,x]}
&& \langle x_{1},\ldots,x_{N}|\psi_{r}(\gamma)\rangle=\det\Big(y_{k}^{-j}(1-y_{k})^{x_{j}}\rme^{\gamma x_{j}}\Big)_{j,k=1,\ldots,N}\\
\label{psiL[y,x]}
&& \langle\psi_{r}(\gamma)|x_{1},\ldots,x_{N}\rangle=\det\Big(y_{k}^{j}(1-y_{k})^{-x_{j}}\rme^{-\gamma x_{j}}\Big)_{j,k=1,\ldots,N}\;.
\end{eqnarray}
These determinants are antisymmetric under the exchange of the $y_{j}$'s, and are thus divisible by the Vandermonde determinant of the $y_{j}$'s. In particular, for the configuration $\mathcal{C}_{X}$ with particles at positions $(X,X+1,\ldots,X+N-1)$, they reduce to
\begin{eqnarray}
\label{psiR[y,X]}
&& \langle\mathcal{C}_{X}|\psi_{r}(\gamma)\rangle=\rme^{\frac{2\rmi\pi p_{r}X}{L}}\rme^{\frac{N(N-1)\gamma}{2}}\bigg(\prod_{j=1}^{N}y_{j}^{-N}\bigg)\!\!\prod_{1\leq j<k\leq N}\!\!\!\!(y_{j}-y_{k})\\
\label{psiL[y,X]}
&& \langle\psi_{r}(\gamma)|\mathcal{C}_{X}\rangle=\rme^{-\frac{2\rmi\pi p_{r}X}{L}}\rme^{-\frac{N(N-1)\gamma}{2}}\bigg(\prod_{j=1}^{N}\frac{y_{j}}{(1-y_{j})^{N-1}}\bigg)\!\!\prod_{1\leq j<k\leq N}\!\!\!\!(y_{k}-y_{j})\;.
\end{eqnarray}

Based on numerical solutions, the expressions above for the eigenvectors and eigenvalues are only valid for generic values of $\gamma$. For specific values of $\gamma$, some eigenstates might be missing. Those can be identified, by adding a small perturbation to $\gamma$, as cases where several $y_{j}$'s coincide, which imply that the determinants in (\ref{psiR[y,x]}) and (\ref{psiL[y,x]}) vanish. This is in particular the case for the stationary eigenstate at $\gamma=0$: in the limit $\gamma\to0$, all $y_{j}$'s converge to $0$ as $\gamma^{1/N}$. This will not be a problem here, as one can always add a small perturbation to $\gamma$ when needed, see also \cite{NW2014.1} for a discussion in XXX and XXZ spin chain.
\end{subsection}

\begin{subsection}{Normalization of Bethe eigenstates}
The eigenvectors (\ref{psiR[y,x]}), (\ref{psiL[y,x]}) are not normalized. In order to write the decomposition of the identity
\begin{equation}
\openone=\sum_{r}\frac{|\psi_{r}(\gamma)\rangle\,\langle\psi_{r}(\gamma)|}{\langle\psi_{r}(\gamma)|\psi_{r}(\gamma)\rangle}\;,
\end{equation}
one needs to compute the scalar products $\langle\psi_{r}(\gamma)|\psi_{r}(\gamma)\rangle$ between left and right eigenstates corresponding to the same Bethe roots (and hence same eigenvalue).

Several results are known, both for on-shell (Bethe roots satisfying the Bethe equations) and off-shell (arbitrary $y_{j}$'s) scalar products. We write explicitly the dependency of the Bethe vectors (\ref{psiR[y,x]}) and (\ref{psiL[y,x]}) on the $y_{j}$'s as $\psi(\vec{y})$. For arbitrary complex numbers $y_{j}$, $w_{j}$, $j=1,\ldots,N$, it was shown \cite{B2009.1,MS2013.1} that
\begin{equation}
\label{norm[y,w] off off}
\langle\psi(\vec{w})|\psi(\vec{y})\rangle=\bigg(\prod_{j=1}^{N}\frac{1-y_{j}}{y_{j}}\,\frac{w_{j}^{N}}{(1-w_{j})^{L}}\bigg)\det\Bigg(\frac{\frac{(1-w_{k})^{L}}{w_{k}^{N-1}}-\frac{(1-y_{j})^{L}}{y_{j}^{N-1}}}{y_{j}-w_{k}}\Bigg)_{j,k=1,\ldots,N}\;.
\end{equation}
For the mixed on-shell / off-shell case, where the $y_{j}$'s verify the Bethe equations while the $w_{j}$'s are arbitrary, one has the Slavnov determinant \cite{S1989.1}
\begin{equation}
\label{norm[y,w] off on}
\langle\psi(\vec{w})|\psi(\vec{y})\rangle=(-1)^{N}\bigg(\prod_{j=1}^{N}\frac{(1-y_{j})^{L+1}}{y_{j}^{N}(1-w_{j})^{L}}\bigg)\bigg(\prod_{j=1}^{N}\prod_{k=1}^{N}(y_{j}-w_{k})\bigg)
\det\Big(\partial_{y_{i}}\mathcal{E}(w_{j},\vec{y})\Big)_{i,j=1,\ldots,N}\;,
\end{equation}
where the derivative with respect to $y_{i}$ is taken before setting the $y_{j}$'s equal to solutions of the Bethe equations. The quantity $\mathcal{E}(\lambda,\vec{y})$ is the eigenvalue of the transfer matrix associated to TASEP with spectral parameter $\lambda$
\begin{equation}
\mathcal{E}(\lambda,\vec{y})=\prod_{j=1}^{N}\frac{1}{1-\lambda^{-1}y_{j}}+\rme^{L\gamma}(1-\lambda)^{L}\prod_{j=1}^{N}\frac{1}{1-\lambda y_{j}^{-1}}\;.
\end{equation}
Finally, for left and right eigenvectors with the same on-shell Bethe roots $y_{j}$ satisfying Bethe equations, the scalar product is equal to the Gaudin determinant \cite{GMCW1981.1,K1982.1}
\begin{equation}
\label{norm[y]}
\langle\psi_{r}(\gamma)|\psi_{r}(\gamma)\rangle=(-1)^{N}\Big(\prod_{j=1}^{N}(1-y_{j})\Big)\\
\det\bigg(\partial_{y_{i}}\log\Big((1-y_{j})^{L}\prod_{k=1}^{N}\frac{y_{k}}{y_{j}}\Big)\bigg)_{i,j=1,\ldots,N}\;.
\end{equation}
Very similar determinantal expressions to (\ref{norm[y,w] off on}) and (\ref{norm[y]}) also exist for more general integrable models, in particular ASEP with particles hopping in both directions. The determinantal formula (\ref{norm[y,w] off off}) for the fully off-shell case seems so far only available for the special case of TASEP.

In this paper, we only consider the on-shell scalar product (\ref{norm[y]}), which can be simplified further by computing the derivative with respect to $y_{i}$ and using the identity
\begin{equation}
\det(\alpha_{i}+\beta_{i}\delta_{i,j})_{i,j=1,\ldots,N}=\bigg(\prod_{j=1}^{N}\beta_{j}\bigg)\bigg(1+\sum_{j=1}^{N}\frac{\alpha_{j}}{\beta_{j}}\bigg)\;,
\end{equation}
where $\delta$ is Kronecker's delta symbol. The scalar product is then equal to
\begin{equation}
\langle\psi_{r}(\gamma)|\psi_{r}(\gamma)\rangle=\frac{L}{N}\bigg(\sum_{j=1}^{N}\frac{y_{j}}{N+(L-N)y_{j}}\bigg)\prod_{j=1}^{N}\Big(L-N+\frac{N}{y_{j}}\Big)\;.
\end{equation}
This normalization is somewhat arbitrary since it depends on the choice of the normalization in the definitions (\ref{psiR[y,x]}), (\ref{psiL[y,x]}). We consider then the configuration $\mathcal{C}_{X}$ with particles at positions $(X,X+1,\ldots,X+N-1)$ and define
\begin{equation}
\mathcal{N}_{r}(\gamma)
=\Omega\,\frac{\langle\mathcal{C}_{X}|\psi_{r}(\gamma)\rangle\langle\psi_{r}(\gamma)|\mathcal{C}_{X}\rangle}{\langle\psi_{r}(\gamma)|\psi_{r}(\gamma)\rangle}\;,
\end{equation}
with $\Omega=\C{L}{N}$ the total number of configurations. One has
\begin{equation}
\label{norm}
\mathcal{N}_{r}(\gamma)
=(-1)^{\frac{N(N-1)}{2}}\rme^{-2\rmi\pi p_{r}(\rho-\frac{1}{L})}
\frac{\rme^{N(N-1)\gamma}\Omega}{N^{N}(\prod_{j=1}^{N}y_{j})^{N-2}}\,\frac{\prod_{j=1}^{N}\prod_{k=j+1}^{N}(y_{j}-y_{k})^{2}}{\Big(\frac{1}{N}\sum_{j=1}^{N}\frac{y_{j}}{\rho+(1-\rho)y_{j}}\Big)\prod_{j=1}^{N}\Big(1+\frac{1-\rho}{\rho}\,y_{j}\Big)}\;.
\end{equation}
Since this formula is based on the rather involved proof of (\ref{norm[y]}) obtained in \cite{K1982.1} for the slightly different case of the XXZ spin chain, we checked it numerically starting from (\ref{psiR[y,x]}), (\ref{psiL[y,x]}) for all systems with $2\leq L\leq10$, $1\leq N\leq L-1$ and all eigenstates. We used the method described in the next section to solve the Bethe equations. Generic values were chosen for the parameter $\gamma$. Perfect agreement was found with (\ref{norm}).

In the basis of configurations, all the elements of the left stationary eigenvector at $\gamma=0$ are equal since $M(0)$ is a stochastic matrix. The same is true for the right stationary eigenvector due to a property of pairwise balance verified by periodic TASEP \cite{SRB1996.1}. Denoting the stationary state by the index $0$, this implies $\mathcal{N}_{0}(0)=1$.

The main goal of this article is the calculation of the asymptotics (\ref{asymptotics norm}) of (\ref{norm}) for large $L$ with fixed density of particles $\rho$ and rescaled fugacity $s$ for the first eigenstates beyond the stationary state.
\end{subsection}

\begin{subsection}{Solution of the Bethe equations}
The Bethe equations of TASEP can be solved in a rather simple way using the fact that they almost decouple, since the right hand side of (\ref{BE[y]}) can be written as $y_{j}^{N}$ times a symmetric function of the $y_{k}$'s independent of $j$. The strategy \cite{GS1992.1,DL1998.1} is then to give a name to that function of the $y_{k}$'s and treat it as a parameter independent of the Bethe roots, than is subsequently fixed using its explicit expression in terms of the $y_{k}$'s. This procedure can be conveniently written \cite{PP2007.1,P2014.1} by introducing the function
\begin{equation}
\label{g}
g:y\mapsto\frac{1-y}{y^{\rho}}\;.
\end{equation}
Indeed, defining the quantity
\begin{equation}
\label{b[y]}
b=\gamma+\frac{1}{L}\,\sum_{j=1}^{N}\log y_{j}\;
\end{equation}
and taking the power $1/L$ of the Bethe equations (\ref{BE[y]}), we observe that there must exist wave numbers $k_{j}$, integers (half-integers) if $N$ is odd (even) such that
\begin{equation}
\label{g(y)}
g(y_{j})=\exp\Big(\frac{2\rmi\pi k_{j}}{L}-b\Big)\;.
\end{equation}
Inverting the function $g$ leads to a rather explicit solution of the Bethe equations as
\begin{equation}
\label{y[k]}
y_{j}=g^{-1}\Big(\exp\Big(\frac{2\rmi\pi k_{j}}{L}-b\Big)\Big)\;.
\end{equation}
This expression is very convenient for large $L$ asymptotic analysis using the Euler-Maclaurin formula.
\end{subsection}

\begin{subsection}{First excited states}
The stationary state corresponds to the choice $k_{j}=k_{j}^{0}$, $j=1,\ldots,N$ with
\begin{equation}
\label{k0}
k_{j}^{0}=j-\frac{N+1}{2}\;.
\end{equation}
This choice closely resembles the Fermi sea of a system of spinless fermions.

We call \textit{first excited states} the (infinitely many) eigenstates of $M(\gamma)$ having a real part scaling as $L^{-3/2}$ in the thermodynamic limit $L\to\infty$ with fixed density of particles $\rho$ and purely imaginary rescaled fugacity $s$. These eigenstates correspond to sets $\{k_{j},j=1,\ldots,N\}$ close to the stationary choice (\ref{k0}). They are built by removing from $\{k_{j}^{0},j=1,\ldots,N\}$ a finite number of $k_{j}$'s located at a finite distance of $\pm N/2$ and adding the same number of $k_{j}$'s at a finite distance of $\pm N/2$ outside of the interval $[-N/2,N/2]$. The first eigenstates are characterized by an equal number of $k_{j}$'s removed and added on each side. In particular, the choice $k_{j}=j-(N-1)/2$ leads to a larger eigenvalue, $\Re E(0)\sim L^{-2/3}$ \cite{P2013.1}, and thus does not belong to the first eigenstates. Numerical checks seem to support the fact that no other choices for the $k_{j}$'s lead to eigenvalues with a real part scaling as $L^{-3/2}$, although a proof of this is missing.

The first excited states can be described by four finite sets of positive half-integers $A_{0}^{\pm},A^{\pm}\subset\mathbb{N}+\tfrac{1}{2}$: the set of $k_{j}$'s removed from (\ref{k0}) are $\{N/2-a,a\in A_{0}^{+}\}$ and $\{-N/2+a,a\in A_{0}^{-}\}$, while the set of $k_{j}$'s added are $\{N/2+a,a\in A^{+}\}$ and $\{-N/2-a,a\in A^{-}\}$, see figure \ref{fig choice k}. The cardinals of the sets verify the constraints
\begin{equation}
\label{m+-}
m_{r}^{+}=|A_{0}^{+}|=|A^{+}|
\quad\text{and}\quad
m_{r}^{-}=|A_{0}^{-}|=|A^{-}|\;.
\end{equation}
We call $m_{r}=m_{r}^{+}+m_{r}^{-}$.

In the following, we use the notation $r$ as a shorthand for $(A_{0}^{+},A^{+},A_{0}^{-},A^{-})$ to refer to the corresponding excited state. The total momentum of an eigenstate, $p_{r}=\sum_{j=1}^{N}k_{j}$, can be written in terms of the four sets as $p_{r}=\sum_{a\in A_{0}^{+}}a+\sum_{a\in A^{+}}a-\sum_{a\in A_{0}^{-}}a-\sum_{a\in A^{-}}a$.

Only the excited states having an eigenvalue with real part scaling as $L^{-3/2}$ contribute to the relaxation for times $T\sim L^{3/2}$: the other eigenstates with larger eigenvalue only give exponentially small corrections when $L\to\infty$. This statement needs however some more justification since, in principle, it could be that the number of higher excited states becomes so large that $TE_{r}(\gamma)$ becomes negligible compared to the "entropy" of the spectrum in the expansion of (\ref{GF[M]}) over the eigenstates. This entropy was studied in \cite{P2013.1} at $\gamma=0$ for the bulk of the spectrum with eigenvalues scaling proportionally to $L$. It was shown that the number of eigenvalues with a real part $-Le$ grows as $\exp(Ls(e))$ with $s(e)\sim e^{2/5}$ for small $e$. Assuming that the $2/5$ exponent still holds for eigenvalues scaling as $L^{-\alpha}$ with $-1<\alpha<3/2$, the contribution to (\ref{GF[M]}) of the entropic part is of order $\exp(L^{(3-2\alpha)/5})$, which is always negligible compared to the contribution of $\rme^{TE_{r}(\gamma)}\sim\exp(L^{3/2-\alpha})$ except at $\alpha=3/2$.
\begin{figure}
  \begin{center}
    \setlength{\unitlength}{0.75mm}
    \begin{picture}(150,70)(0,-15)
      \put(0,40){\color{lightblue}\polygon*(40,0)(110,0)(110,5)(40,5)}
      \put(0,40){\line(1,0){150}}
      \put(0,45){\line(1,0){150}}
      \multiput(10,40)(5,0){13}{\line(0,1){5}}
      \multiput(140,40)(-5,0){13}{\line(0,1){5}}
      \put(2,42){$\ldots$}
      \put(73,42){$\ldots$}
      \put(143,42){$\ldots$}
      \put(20,10){\color{lightblue}\polygon*(0,0)(5,0)(5,5)(0,5)}
      \put(35,10){\color{lightblue}\polygon*(0,0)(5,0)(5,5)(0,5)}
      \put(45,10){\color{lightblue}\polygon*(0,0)(5,0)(5,5)(0,5)}
      \put(0,10){\color{lightblue}\polygon*(55,0)(90,0)(90,5)(55,5)}
      \put(100,10){\color{lightblue}\polygon*(0,0)(5,0)(5,5)(0,5)}
      \put(115,10){\color{lightblue}\polygon*(0,0)(5,0)(5,5)(0,5)}
      \put(120,10){\color{lightblue}\polygon*(0,0)(5,0)(5,5)(0,5)}
      \put(130,10){\color{lightblue}\polygon*(0,0)(5,0)(5,5)(0,5)}
      \put(0,10){\line(1,0){150}}
      \put(0,15){\line(1,0){150}}
      \multiput(10,10)(5,0){13}{\line(0,1){5}}
      \multiput(140,10)(-5,0){13}{\line(0,1){5}}
      \put(2,12){$\ldots$}
      \put(73,12){$\ldots$}
      \put(143,12){$\ldots$}
      \put(42.5,38){\vector(-0.9,-1){20}}
      \put(52.5,38){\vector(-0.65,-1){14.5}}
      \put(92.5,38){\vector(1.1,-1){24}}
      \put(97.5,38){\vector(1.1,-1){24}}
      \put(107.5,38){\vector(1.1,-1){24}}
      \put(21.4,4.5){$\frac{7}{2}$}
      \put(36.4,4.5){$\frac{1}{2}$}
      \put(41.4,4.5){$\frac{1}{2}$}
      \put(51.4,4.5){$\frac{5}{2}$}
      \put(91.4,4.5){$\frac{7}{2}$}
      \put(96.4,4.5){$\frac{5}{2}$}
      \put(106.4,4.5){$\frac{1}{2}$}
      \put(116.4,4.5){$\frac{3}{2}$}
      \put(121.4,4.5){$\frac{5}{2}$}
      \put(131.4,4.5){$\frac{9}{2}$}
      \put(21,2){$\underbrace{\hspace{18\unitlength}}$}
      \put(41,2){$\underbrace{\hspace{13\unitlength}}$}
      \put(91,2){$\underbrace{\hspace{18\unitlength}}$}
      \put(111,2){$\underbrace{\hspace{23\unitlength}}$}
      \put(30,-4){$A^{-}$}
      \put(47.5,-4){$A_{0}^{-}$}
      \put(100,-4){$A_{0}^{+}$}
      \put(122.5,-4){$A^{+}$}
      \put(75,50){\vector(1,0){35}}
      \put(75,50){\vector(-1,0){35}}
      \put(72,52){$\sim L$}
      \put(40,-10){\vector(1,0){15}}
      \put(40,-10){\vector(-1,0){20}}
      \put(35,-15){$\sim L^{0}$}
      \put(110,-10){\vector(1,0){25}}
      \put(110,-10){\vector(-1,0){20}}
      \put(107,-15){$\sim L^{0}$}
    \end{picture}
  \end{center}
  \caption{Representation of the choices of the numbers $k_{j}$, $j=1,\ldots,N$ characterizing the first eigenstates. The red squares represent the $k_{j}$'s chosen. The upper line corresponds to the choice for the stationary state (\ref{k0}). The lower line corresponds to a generic eigenstate close to the stationary state, with excitations characterized by sets $A_{0}^{-}=\{\frac{1}{2},\frac{5}{2}\}$, $A^{-}=\{\frac{1}{2},\frac{7}{2}\}$, $A_{0}^{+}=\{\frac{1}{2},\frac{5}{2},\frac{7}{2}\}$, $A^{+}=\{\frac{3}{2},\frac{5}{2},\frac{9}{2}\}$ of cardinals $m_{r}^{-}=|A_{0}^{-}|=|A^{-}|=2$ and $m_{r}^{+}=|A_{0}^{+}|=|A^{+}|=3$.}
  \label{fig choice k}
\end{figure}
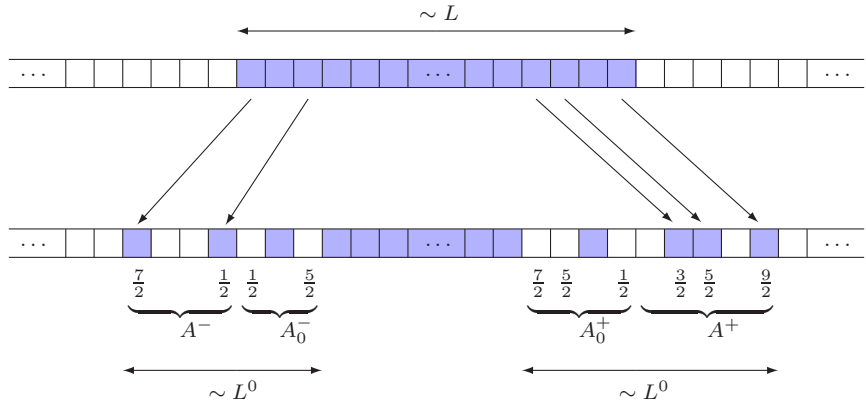
\end{subsection}

\begin{subsection}{Field \texorpdfstring{$\varphi_{r}$}{phi}}
The eigenvalue corresponding to the eigenstate $r$ can be nicely written in terms of a function $\eta_{r}$ \cite{P2014.1}. From the result stated in section \ref{section expansion norm} about the normalization of Bethe states, the function $\varphi_{r}(u)=-(2\pi)^{-3/2}\eta_{r}'(\tfrac{u}{2\pi})$ seems in fact the "good" object to describe the first excited states. It is defined by
\begin{eqnarray}
\label{phi[A,zeta]}
&& \varphi_{r}(u)=2\sqrt{\pi}\Big(\rme^{\rmi\pi/4}\zeta\big(-\frac{1}{2},\frac{1}{2}+\frac{\rmi u}{2\pi}\big)+\rme^{-\rmi\pi/4}\zeta\big(-\frac{1}{2},\frac{1}{2}-\frac{\rmi u}{2\pi}\big)\Big)\\
&&\hspace{13mm} +\rmi\sqrt{2}\Big(\sum_{a\in A_{0}^{+}}\sqrt{u+2\rmi\pi a}+\sum_{a\in A^{-}}\sqrt{u+2\rmi\pi a}
-\sum_{a\in A_{0}^{-}}\sqrt{u-2\rmi\pi a}-\sum_{a\in A^{+}}\sqrt{u-2\rmi\pi a}\Big)\;.\nonumber
\end{eqnarray}
The Hurwitz zeta function (\ref{Hurwitz zeta}) can be seen as a kind of renormalization of an infinite contribution of the Fermi sea to $\varphi_{r}$. Indeed, using (\ref{sum_power[zeta]}), (\ref{Hurwitz zeta asymptotics}) and introducing the quantities $\chi_{a}^{\pm}(u)=\pm\rmi\sqrt{2u\pm4\rmi\pi a}$, we observe that $\varphi_{r}$ can be written as a sum over momenta of elementary excitations $k_{j}$ near $\pm N/2$ as
\begin{equation}
\varphi_{r}(u)=\lim_{M\to\infty}\Big(-\frac{4\sqrt{2\pi}}{3}M^{3/2}+\frac{\sqrt{2}u}{\sqrt{\pi}}\sqrt{M}+\sum_{a\in B_{M}^{+}}\chi_{a}^{-}(u)+\sum_{a\in B_{M}^{-}}\chi_{a}^{+}(u)\Big)\;,
\end{equation}
with $B_{M}^{\pm}=\big(\{-M+\half,-M+\tfrac{3}{2},\ldots,-\half\}\backslash(-A_{0}^{\pm})\big)\cup A^{\pm}$.

The $\zeta$ functions are responsible for branch points $\pm\rmi\pi$ for the function $\varphi_{r}$. The square roots provide additional branch points in $\pm2\rmi\pi(\mathbb{N}+\half)$. We define in the following $\varphi_{r}$ with branch cuts $[\rmi\pi,\rmi\infty)$ and $(-\rmi\infty,-\rmi\pi]$.

Using the relation between polylogarithms and Hurwitz zeta function, the field for the stationary state can be written as
\begin{equation}
\label{phi0[Li]}
\varphi_{0}(u)=-\frac{1}{\sqrt{2\pi}}\,\Li_{3/2}(-\rme^{u})\;,
\end{equation}
which is valid for $\Re u<0$, and for $\Re u>0$ with $|\Im u|<\pi$.

Our main result (\ref{asymptotics norm}), which expresses the asymptotics of the norm of Bethe eigenstates in terms of the free action of $\varphi_{r}$, seems to indicate that $\varphi_{r}$ should be interpreted as a field, whose physical meaning is unclear at the moment.
\end{subsection}

\begin{subsection}{Large \texorpdfstring{$L$}{L} expansion of the parameter \texorpdfstring{$b$}{b}}
\label{section expansion b}
For all first excited states, the quantity $b$ converges in the thermodynamic limit to $b_{0}$ \cite{P2014.1}, equal to
\begin{equation}
\label{b0}
b_{0}=\rho\log\rho+(1-\rho)\log(1-\rho)\;.
\end{equation}
Writing the correction as
\begin{equation}
\label{b[c]}
b=b_{0}+\frac{2\pi c}{L}\;,
\end{equation}
a small generalization of \cite{P2014.1} to non-zero rescaled fugacity $s$ leads to the large $L$ expansion
\begin{equation}
\label{phi(c)}
\varphi_{r}(2\pi c)\simeq s-\frac{2\rmi\pi(1-2\rho)p_{r}}{3\sqrt{\rho(1-\rho)}\sqrt{L}}\;.
\end{equation}
This expansion follows from applying the Euler-Maclaurin formula to (\ref{b[y]}), with Bethe roots $y_{j}$ given by (\ref{y[k]}).
\end{subsection}

\begin{subsection}{Large \texorpdfstring{$L$}{L} expansion of the eigenvalues}
\label{section expansion E}
Another small extension of \cite{P2014.1} to nonzero rescaled fugacity $s$ gives the expansion up to order $L^{-3/2}$ of the eigenvalue $E_{r}(\gamma)$ as
\begin{equation}
\label{asymptotics E}
E_{r}(\gamma)\simeq\frac{s\sqrt{\rho(1-\rho)}}{\sqrt{L}}-\frac{2\rmi\pi(1-2\rho)p_{r}}{L}
+\frac{\sqrt{\rho(1-\rho)}}{L^{3/2}}\lim_{\Lambda\to\infty}\Big(D^{\Lambda}+\int_{-\Lambda}^{2\pi c}\rmd u\,\varphi_{r}(u)\Big)\;,
\end{equation}
where
\begin{equation}
\label{DLambda}
D^{\Lambda}=\frac{4\sqrt{2}m_{r}}{3}\,\Lambda^{3/2}-2\sqrt{2}\rmi\pi\Big(\sum_{a\in A_{0}^{+}}a+\sum_{a\in A^{-}}a-\sum_{a\in A_{0}^{-}}a-\sum_{a\in A^{+}}a\Big)\sqrt{\Lambda}\;
\end{equation}
cancels the divergent contribution from the integral. This expansion follows from the application of the Euler-Maclaurin formula to (\ref{E[y]}).
\end{subsection}

\begin{subsection}{Large \texorpdfstring{$L$}{L} expansion of the norm of Bethe eigenstates}
\label{section expansion norm}
In section \ref{section asymptotics}, we derive the large $L$ asymptotics (\ref{asymptotics (Xi3*Xi4)/(Xi1*Xi2)}) for the normalization of Bethe states, with $b$ written as (\ref{b[c]}) and $c$ arbitrary. Writing the solution of (\ref{phi(c)}) at leading order in $L$ as $2\pi c=\varphi_{r}^{-1}(s)$, the asymptotics of the normalization of Bethe states is obtained as
\begin{eqnarray}
\label{asymptotics norm}
&& \mathcal{N}_{r}(\gamma)\simeq\rme^{-2\rmi\pi\rho p_{r}}\,\rme^{-s\sqrt{\rho(1-\rho)}\sqrt{L}}\nonumber\\
&&\hspace{13mm} \times\frac{(\pi^{2}/4)^{m_{r}^{2}}}{(-4\pi^{2})^{m_{r}}}\,\omega(A_{0}^{+})^{2}\omega(A_{0}^{-})^{2}\omega(A^{+})^{2}\omega(A^{-})^{2}\omega(A_{0}^{+},A_{0}^{-})^{2}\omega(A^{+},A^{-})^{2}\\
&&\hspace{13mm} \times\frac{\rme^{\varphi_{r}^{-1}(s)}}{\sqrt{2\pi}\,\varphi_{r}'(\varphi_{r}^{-1}(s))}\,
\lim_{\Lambda\to\infty}\exp\Big(-2m_{r}^{2}\log\Lambda+\int_{-\Lambda}^{\varphi_{r}^{-1}(s)}\rmd u\,(\varphi_{r}'(u))^{2}\Big)\;,\nonumber
\end{eqnarray}
with the combinatorial factors
\begin{equation}
\label{omega(A)}
\omega(A)=\prod_{\substack{a,a'\in A\\a<a'}}(a-a')
\quad\text{and}\quad
\omega(A,A')=\prod_{a\in A}\prod_{a'\in A'}(a+a')\;.
\end{equation}
This is the main technical result of the paper. The field $\varphi_{r}$ is defined by (\ref{phi[A,zeta]}).

One recovers the stationary value $\mathcal{N}_{0}(0)=1$ using the fact that the solution $c$ of $\varphi_{0}(2\pi c)=s$ for the stationary state goes  to $-\infty$ when $s$ goes to $0$ and the expression (\ref{phi0[Li]}) of $\varphi_{0}$ as a polylogarithm. For technical reasons, our derivation of (\ref{asymptotics norm}) requires $\Re c>0$ for all the other eigenstates. This implies that $\Re s$ can not be too small. Numerical resolution of $\varphi_{r}(2\pi c)=s$ for the first eigenstates seem to indicate that the condition $\Re s\geq0$ is always sufficient.
\end{subsection}

\begin{subsection}{Numerical checks of the asymptotic expansion}
\label{section numerics}
Bulirsch-Stoer (BST) algorithm (see \textit{e.g.} \cite{HS1988.1}) is an extrapolation method for convergence acceleration of algebraically converging sequences $q_{L}$, $L\in\mathbb{N}^{*}$. It assumes that $q_{L}$ behaves for large $L$ as $p_{0}+p_{1}L^{-\omega}+p_{2}L^{-2\omega}+\ldots$ for some exponent $\omega>0$. Given values $q_{j}$, $j=1,\ldots,M$ of the sequence, the algorithm provides an estimation of the limit $p_{0}$, together with an estimation of the error. In the usual case where one does not know the value of the parameter $\omega$, it has to be estimated by trying to minimize the estimation of the error, which requires some educated guesswork. In the case considered in this paper, however, we know that the norm of the eigenstates has an asymptotic expansions in $1/\sqrt{L}$. One can then set from the beginning $\omega=1/2$ when applying BST algorithm. The convergence of the estimation of $p_{0}$ to its exact value is then exponentially fast in the number $M$ of values of the sequence supplied to the algorithm, although the larger $M$ is, the more precision is needed for the $q_{j}$'s due to fast propagation of rounding errors. BST algorithm thus allows to check to very high accuracy the asymptotics obtained.

We used BST algorithm in order to check (\ref{asymptotics norm}) for all $139$ first eigenstates with $\sum_{a\in A_{0}^{+}}a+\sum_{a\in A^{+}}a+\sum_{a\in A_{0}^{-}}a+\sum_{a\in A^{-}}a\leq6$ (giving $57$ different values for the norms due to degeneracies). We computed numerically the exact formula (\ref{norm}) in terms of the Bethe roots, divided by all the factors of the asymptotics except the exponential of the integral, for several values of the system size $L$ and fixed density of particles $\rho$. We then compared the result of BST algorithm for each value of $\rho$ with the numerical value of the exponential of the integral, which was computed by cutting the integral into three pieces: from $-\infty$ to $-1$ with the integrand $\varphi_{r}'(u)^{2}+2m_{r}^{2}/u$, from $-1$ to $0$ with the integrand $\varphi_{r}'(u)^{2}$ and from $0$ to $\varphi_{r}^{-1}(s)$ with the integrand $\varphi_{r}'(u)^{2}$. The first piece absorbs the divergence at $u\to-\infty$, while the last two pieces make sure that the path of integration does not cross the branch cuts of $\varphi_{r}$.

All the computations were done with the generic value $s=0.2+\rmi$ for the rescaled fugacity. Solving numerically the equation (\ref{b[y]}) for $b$, calculating the Bethe roots from (\ref{y[k]}), inverting the field $\varphi_{r}$, and evaluating numerically the integrals are relatively costly in computer time, especially with a large number of digits. The exact formula was computed with $200$ significant digits, for $\rho=1/2$ with $L=12,14,16,\ldots,200$, for $\rho=1/3$ with $L=18,21,24,\ldots,300$, for $\rho=1/4$ with $L=24,28,32,\ldots,400$, and for $\rho=1/5$ with $L=30,35,40,\ldots,500$. The estimated relative error from BST algorithm was lower than $10^{-50}$ in all cases. Comparing with the numerical evaluation of the integrals, we found a perfect agreement within at least $50$ digits in relative accuracy.
\end{subsection}

\begin{subsection}{Current fluctuations}
From the asymptotics (\ref{asymptotics norm}) of the norm, we are now in position to write the large $L$, $T$ limit with
\begin{equation}
T=\frac{t\,L^{3/2}}{\sqrt{\rho(1-\rho)}}\;
\end{equation}
of the generating function for the current fluctuations, in the case of an evolution conditioned on the initial condition $\mathcal{C}_{X}$ with particles at position $(X,X+1,\ldots,X+N-1)$ and on the final condition $\mathcal{C}_{Y}$ with particles at position $(Y,Y+1,\ldots,Y+N-1)$. The distance between $X$ and $Y$ is taken as
\begin{equation}
\label{Y-X}
Y-X=(1-2\rho)T+(x+\rho)L\;.
\end{equation}
The first term comes from the term of order $L^{-1}$ in the eigenvalue (\ref{asymptotics E}) and corresponds to the group velocity, while the second term defines a rescaled distance $x$ on the ring. The current fluctuations are then defined as
\begin{equation}
\xi_{t,x}=\frac{Q_{T}-JLT}{\sqrt{\rho(1-\rho)}L^{3/2}}\;,
\end{equation}
where the mean value of the current per site $J$ has to be set equal to
\begin{equation}
\label{J}
JLT=\rho(1-\rho)LT-\rho(1-\rho)L^{2}\;.
\end{equation}
Its leading term $\rho(1-\rho)$ is the stationary value of the current. The negative correction $-\rho(1-\rho)L^{2}$ is caused by the beginning and the end of the evolution, during which the particles can not easily move due to the initial and final conditions chosen. Its value follows \cite{P2015.1} from Burgers' equation.

Multiplying the norm (\ref{asymptotics norm}) by $\rme^{2\rmi\pi p_{r}(Y-X)/L}$ to change the final state from $\mathcal{C}_{X}$ to $\mathcal{C}_{Y}$, the choices (\ref{Y-X}) and (\ref{J}) imply for the generating function of the current fluctuations $G_{t,x}(s)=\langle\rme^{\gamma(Q_{T}-JLT)}\rangle=\langle\rme^{s\xi_{t,x}}\rangle$ the large $L$, $T$ limit
\begin{equation}
G_{t,x}(s)\to\frac{1}{Z_{t}}\sum_{r}\omega_{r}^{2}\,\frac{\rme^{2\rmi\pi p_{r}x}\,\rme^{\varphi_{r}^{-1}(s)}}{\varphi_{r}'(\varphi_{r}^{-1}(s))}\,\exp\Bigg[\lim_{\Lambda\to\infty} R^{\Lambda}+\int_{-\Lambda}^{\varphi_{r}^{-1}(s)}\rmd u\,\Big(\varphi_{r}'(u)^{2}+t\,\varphi_{r}(u)\Big)\Bigg]\;.
\end{equation}
The sum is over the infinitely many first excited states $r$ of $M(\gamma)$, characterized by the four sets of positive half-integers $A_{0}^{+}$, $A_{0}^{-}$, $A^{+}$, $A^{-}$ with the equalities on their number of elements $m_{r}^{+}=|A_{0}^{+}|=|A^{+}|$, $m_{r}^{-}=|A_{0}^{-}|=|A^{-}|$. The function $\varphi_{r}$ is defined in (\ref{phi[A,zeta]}). The normalization $Z_{t}$ is such that the generating function is equal to $1$ when the rescaled fugacity $s=0$. The regularization of the integral at $u\to-\infty$ is $R^{\Lambda}=tD^{\Lambda}-2m_{r}^{2}\log\Lambda$ with $D^{\Lambda}$ given by (\ref{DLambda}). The combinatorial factor $\omega_{r}$ is equal to
\begin{equation}
\omega_{r}^{2}=\frac{(-1)^{m_{r}}(\pi^{2}/4)^{m_{r}^{2}}}{(4\pi^{2})^{m_{r}}}\,
\omega(A_{0}^{+})^{2}\omega(A_{0}^{-})^{2}\omega(A^{+})^{2}\omega(A^{-})^{2}
\omega(A_{0}^{+},A_{0}^{-})^{2}\omega(A^{+},A^{-})^{2}\;,
\end{equation}
with factors defined in (\ref{omega(A)}) and $m_{r}=m_{r}^{+}+m_{r}^{-}$.

Taking $s$ purely imaginary inside the generating function, the probability distribution $P_{\xi}$ of the random variable $\xi_{t,x}$ can be extracted by Fourier transform
\begin{equation}
P_{\xi}(w)=\int_{-\infty}^{\infty}\frac{\rmd s}{2\pi}\,\rme^{\rmi sw}G_{t,x}(-\rmi s)\;.
\end{equation}
We observe that the integral over $s$ can be nicely replaced by an integral over $d=\varphi_{r}^{-1}(s)$ on some curve in the complex plane. The Jacobian of this change of variables precisely cancels the denominator $\varphi_{r}'(\varphi_{r}^{-1}(s))$ in the generating function.
\end{subsection}
\end{section}

\begin{section}{Euler-Maclaurin formula}
\label{section Euler-Maclaurin}
In order to obtain the large $L$ limit for the normalization of the eigenstates, we need to compute the asymptotics of various sums (and products) with a summation range growing as $L$ and a summand involving the summation index $j$ as $j/L$. Such asymptotics can be performed using the Euler-Maclaurin formula \cite{H1949.1}. It turns out that the sums considered here have various singularities (square root, logarithm and worse) at both ends of the summation range, for which the most naive version of the Euler-Maclaurin formula (\ref{EM regular}) does not work. We discuss here some adaptations of the Euler-Maclaurin formula to logarithmic singularities (Stirling's formula), non-integer powers (Hurwitz zeta function), and logarithm of a difference of square roots ($\sqrt{\text{Stirling}}$ formula). We begin with one-dimensional sums, and consider then sums on some two dimensional domains necessary to treat the Vandermonde determinant of Bethe roots in (\ref{norm}).

\begin{subsection}{One-dimensional sums}

\begin{subsubsection}{Functions without singularities}
The Euler-Maclaurin formula gives an asymptotic expansion for the difference between a Riemann sum and the corresponding integral. Let $M$ and $L$ be positive integers, $\mu=M/L$ their ratio, and $f$ a function with no singularities in a region which contains the segment $[0,\mu]$. Then, for large $M$, $L$ with fixed $\mu$, the Euler-Maclaurin formula states that
\begin{equation}
\label{EM regular}
\sum_{j=1}^{M}f\Big(\frac{j+d}{L}\Big)\simeq L\Big(\int_{0}^{\mu}\rmd u\,f(u)\Big)+(\mathcal{R}_{L}f)[\mu,d]-(\mathcal{R}_{L}f)[0,d]\;,
\end{equation}
where the remainder term is expressed in terms of the Bernoulli polynomials $B_{\ell}$ as
\begin{equation}
\label{EM remainder term}
(\mathcal{R}_{L}f)[\mu,d]=\sum_{\ell=1}^{\infty}\frac{B_{\ell}(d+1)f^{(\ell-1)}(\mu)}{\ell!L^{\ell-1}}\;.
\end{equation}

A simple derivation of (\ref{EM regular}) using Hurwitz zeta function $\zeta$ (\ref{Hurwitz zeta}) consists in replacing the function $f$ by its Taylor series at $0$ in the sum. In order to perform the summation over $j$ at each order in the Taylor series, we use
\begin{equation}
\label{sum_power[zeta]}
\sum_{j=1}^{M}(j+d)^{\nu}=\zeta(-\nu,d+1)-\zeta(-\nu,M+d+1)\;,
\end{equation}
which follows directly from the definition (\ref{Hurwitz zeta}) for $\nu<-1$, and then for all $\nu\neq-1$ by analytic continuation. It gives
\begin{equation}
\sum_{j=1}^{M}f\Big(\frac{j+d}{L}\Big)=\sum_{k=0}^{\infty}f^{(k)}(0)\,\frac{\zeta(-k,d+1)-\zeta(-k,M+d+1)}{k!L^{k}}\;.
\end{equation}
The $\zeta$ function whose argument depends on $M$ can be expanded for large $M=\mu L$ using (\ref{Hurwitz zeta asymptotics}). This leads to
\begin{equation}
\sum_{j=1}^{M}f\Big(\frac{j+d}{L}\Big)
\simeq\sum_{k=0}^{\infty}f^{(k)}(0)\,\frac{\zeta(-k,d+1)}{k!L^{k}}
+L\sum_{k=0}^{\infty}f^{(k)}(0)\,\frac{\mu^{k+1}}{(k+1)!}
-\sum_{r=0}^{\infty}\frac{\zeta(-r,d+1)}{r!L^{r}}\sum_{k=0}^{\infty}f^{(k+r)}(0)\,\frac{\mu^{k}}{k!}\;.
\end{equation}
The second term on the right is the Taylor series at $\mu=0$ of the integral from $0$ to $\mu$ of $f$, while in the last term, we recognize the Taylor expansion at $\mu=0$ of $f^{(r)}(\mu)$. Expressing the remaining $\zeta$ function in terms of Bernoulli polynomials using (\ref{Bernoulli[zeta]}), we arrive at (\ref{EM regular}).
\end{subsubsection}

\begin{subsubsection}{Logarithmic singularity: Stirling's formula}
For functions having a singularity at the origin, (\ref{EM regular}) can no longer be used, since the derivatives at $0$ of the function become infinite. A well known example is Stirling's formula for the $\Gamma$ function, for which one has a logarithmic singularity. One has the identity
\begin{equation}
\label{sum_log[Gamma]}
\sum_{j=1}^{M}\log(j+d)=\log\Gamma(M+d+1)-\log\Gamma(d+1)\;,
\end{equation}
valid for $d+1\not\in\mathbb{R}^{-}$, where the log $\Gamma$ function $\log\Gamma(z)$ is defined as the analytic continuation with $\log\Gamma(1)=0$ of $\log(\Gamma(z))$ to $\mathbb{C}$ minus the branch cut $\mathbb{R}^{-}$. Then, Stirling's formula can be stated as the asymptotic expansion
\begin{equation}
\label{EM Stirling}
\sum_{j=1}^{M}\log\Big(\frac{j+d}{L}\Big)
\simeq L\Big(\int_{0}^{\mu}\rmd u\,\log u\Big)+(\mathcal{R}_{L}\log)[\mu,d]+(d+\half)\log L+\log\sqrt{2\pi}-\log\Gamma(d+1)\;.
\end{equation}
This has the same form as the Euler-Maclaurin formula for a regular function (\ref{EM regular}), except for the remainder term at $0$, which involves the non trivial constant $\log\sqrt{2\pi}$ when $d=0$. The constant term is analytic in $d$ except for the branch cut of $\log\Gamma$ if $\log\Gamma$ is interpreted as the log $\Gamma$ function and not the logarithm of the $\Gamma$ function. We use this prescription in the rest of the paper.
\end{subsubsection}

\begin{subsubsection}{Non-integer power singularity: Hurwitz zeta function}
Another example of singularities is non-integer power functions. Using (\ref{sum_power[zeta]}) for $\nu\neq-1$ and
\begin{equation}
\label{sum_inverse[Gamma]}
\sum_{j=1}^{M}\frac{1}{j+d}=-\frac{\Gamma'(d+1)}{\Gamma(d+1)}+\frac{\Gamma'(M+d+1)}{\Gamma(M+d+1)}\;
\end{equation}
for $\nu=-1$, the asymptotics of $\zeta$ (\ref{Hurwitz zeta asymptotics}) and of $\Gamma$ give
\begin{equation}
\label{EM power}
\sum_{j=1}^{M}\Big(\frac{j+d}{L}\Big)^{\nu}
\simeq L\Big(\mint_{0}^{\mu}\rmd u\,u^{\nu}\Big)+(\mathcal{R}_{L}(\cdot)^{\nu})[\mu,d]
+\left\{\begin{array}{cll}
  \frac{\zeta(-\nu,d+1)}{L^{\nu}}
  & \quad & \nu\neq-1\\
  -L\,\frac{\Gamma'(d+1)}{\Gamma(d+1)}
  & \quad & \nu=-1
\end{array}\right.\;,
\end{equation}
where $(\cdot)^{\nu}$ denotes the function $x\mapsto x^{\nu}$. The modified integral is equal to
\begin{equation}
\label{mint}
\mint_{0}^{\mu}\rmd u\,u^{\nu}=
\left\{\begin{array}{cll}
  \frac{\mu^{\nu+1}}{\nu+1}
  & \quad & \nu\neq-1\\
  \log\mu-\log L^{-1}
  & \quad & \nu=-1
\end{array}\right.\;.
\end{equation}
It is equal to the usual, convergent, definition of the integral only in the case $\nu>-1$. Both cases in (\ref{EM power}) can be unified by replacing $\zeta$ by $\tilde{\zeta}$ defined in (\ref{Hurwitz zeta tilde}).
\end{subsubsection}

\begin{subsubsection}{Logarithm of a sum of two square roots: \texorpdfstring{$\sqrt{\text{Stirling}}$}{sqrt(Stirling)} formula}
In the calculation of the asymptotics of the normalization of Bethe states, more complicated singularities appear, with functions that depend themselves on $L$. We define
\begin{equation}
\label{alpha+-}
\alpha_{\pm}(u,q)=\log(\sqrt{u}\pm\sqrt{q})\;.
\end{equation}
One has the asymptotic expansion
\begin{eqnarray}
\label{EM sqrt(Stirling)}
&& \sum_{j=1}^{M}\alpha_{\pm}\Big(\frac{j+d}{L},\frac{q}{L}\Big)
\simeq L\Big(\int_{0}^{\mu}\rmd u\,\alpha_{\pm}(u,q/L)\Big)
+(\mathcal{R}_{L}\alpha_{\pm}(\cdot,q/L))[\mu,d]
+\frac{q}{2}-\frac{q\log q}{2}\\
&&\hspace{32mm} +\frac{\rmi\pi(1\mp1)q}{2}\,\sgn(\arg q)
+\frac{(d+\frac{1}{2})\log L}{2}+\frac{\log(2\pi)}{4}-\frac{\log\Gamma(d-q+1)}{2}\nonumber\\
&&\hspace{32mm} \pm\int_{0}^{q}\rmd u\,\frac{\zeta(\frac{1}{2},d+u-q+1)}{2\sqrt{u}}\;,\nonumber
\end{eqnarray}
where the first integral is equal to
\begin{equation}
\int_{0}^{\mu}\rmd u\,\alpha_{\pm}(u,q)=-\frac{\mu}{2}\pm\sqrt{\mu}\sqrt{q}+\frac{q\log q}{2}+\frac{\rmi\pi(-1\pm1)q}{2}\,\sgn(\arg q)+(\mu-q)\log(\sqrt{\mu}\pm\sqrt{q})\;.
\end{equation}
The argument of $q$ is taken in the interval $(-\pi,\pi)$. The path of integration for the second integral in (\ref{EM sqrt(Stirling)}) is required to avoid the branch cuts of the integrand coming from the square root and the $\zeta$ function.

The asymptotic expansion (\ref{EM sqrt(Stirling)}) is a kind of square root version of Stirling's formula since $\alpha_{+}(u,q)+\alpha^{-}(u,q)=\log(u-q)$. Indeed, adding (\ref{EM sqrt(Stirling)}) for $\alpha_{+}$ and $\alpha_{-}$ gives (\ref{EM Stirling}) after using the property
\begin{equation}
\label{EM remainder add}
\Big(\mathcal{R}_{L}f\Big(\cdot-\frac{q}{L}\Big)\Big)\big[\mu,d\big]=(\mathcal{R}_{L}f)\big[\mu-\frac{q}{L},d\big]=(\mathcal{R}_{L}f)[\mu,d-q]+L \int_{\mu-\frac{q}{L}}^{\mu}\!\!\!\rmd u\,f(u)\;,
\end{equation}
which follows from the relation (\ref{Bernoulli sum}) satisfied by the Bernoulli polynomials.

The expansion (\ref{EM sqrt(Stirling)}) is a bit more complicated to show than (\ref{EM Stirling}) or (\ref{EM power}). It can be derived by using the summation formula
\begin{eqnarray}
\label{sum_log_sqrt[Gamma,zeta]}
&& \sum_{j=1}^{M}\log(\sqrt{j+d}\pm\sqrt{q})=
\frac{\log\Gamma(M+d-q+1)}{2}-\frac{\log\Gamma(d-q+1)}{2}\\
&&\hspace{40mm} \pm\int_{0}^{q}\rmd u\,\frac{\zeta(\frac{1}{2},d-q+u+1)-\zeta(\frac{1}{2},M+d-q+u+1)}{2\sqrt{u}}\;,\nonumber
\end{eqnarray}
which can be proved starting from the identity
\begin{equation}
\label{d log(sqrt-sqrt)}
\partial_{\lambda}\log(\sqrt{j+d+\lambda}\pm\sqrt{q+\lambda})=\pm\frac{1}{2\sqrt{j+d+\lambda}\,\sqrt{q+\lambda}}\;.
\end{equation}
Indeed, summing (\ref{d log(sqrt-sqrt)}) over $j$ using (\ref{sum_power[zeta]}) and integrating over $\lambda$, there exist a quantity $K_{M}(d,q)$, independent of $\lambda$, such that
\begin{equation}
\sum_{j=1}^{M}\log(\sqrt{j+d+\lambda}\pm\sqrt{q+\lambda})
=K_{M}(d,q)\pm\int_{0}^{\lambda}\rmd u\,\frac{\zeta(\frac{1}{2},d+u+1)-\zeta(\frac{1}{2},M+d+u+1)}{2\sqrt{q+u}}\;.
\end{equation}
The constant of integration can be fixed from the special case $\lambda=-q$, using (\ref{sum_log[Gamma]}). Taking $\lambda=0$ in the previous equation finally gives (\ref{sum_log_sqrt[Gamma,zeta]}).

The asymptotic expansion (\ref{EM sqrt(Stirling)}) is a consequence of the summation formula (\ref{sum_log_sqrt[Gamma,zeta]}). Using (\ref{sum_power[zeta]}) and (\ref{sum_log[Gamma]}) we rewrite (\ref{sum_log_sqrt[Gamma,zeta]}) as
\begin{equation}
\sum_{j=1}^{M}\log(\sqrt{j+d}\pm\sqrt{q})=
\frac{1}{2}\,\sum_{j=1}^{M}\log(j+d-q)
\pm\int_{0}^{q}\frac{\rmd u}{2\sqrt{u}}\,\sum_{j=1}^{M}\frac{1}{\sqrt{j+d-q+u}}\;,
\end{equation}
where the integration is on a contour that avoids the branch cuts of the square roots. The asymptotic expansions (\ref{EM Stirling}) and (\ref{EM power}) give
\begin{eqnarray}
&& \sum_{j=1}^{M}\log\Big(\sqrt{\frac{j+d}{L}}\pm\sqrt{\frac{q}{L}}\Big)\simeq
\frac{L}{2}\,\Big(\int_{0}^{\mu}\rmd u\,\log u\Big)
+\half(d-q+\half)\log L+\frac{\log(2\pi)}{4}-\frac{\log\Gamma(d-q+1)}{2}\nonumber\\
&&\hspace{42mm}
\pm\int_{0}^{q}\rmd u\,\frac{\zeta(\tfrac{1}{2},d-q+u+1)}{2\sqrt{u}}
\pm\sqrt{L}\Big(\int_{0}^{q}\frac{\rmd u}{2\sqrt{u}}\Big)\Big(\int_{0}^{\mu}\frac{\rmd v}{\sqrt{v}}\Big)\\
&&\hspace{42mm}
\pm\frac{1}{2\sqrt{L}}\int_{0}^{q}\frac{\rmd u}{\sqrt{u}}\Big(\mathcal{R}_{L}\frac{1}{\sqrt{\cdot}}\Big)[\mu,d-q+u]+\frac{(\mathcal{R}_{L}\log)[\mu,d-q]}{2}\;.\nonumber
\end{eqnarray}
The operator $\mathcal{R}_{L}$, defined in (\ref{EM remainder term}), is linear. From (\ref{EM remainder add}), it verifies
\begin{equation}
\int_{0}^{q}\rmd u\,h(u)\,(\mathcal{R}_{L}f)[\mu,d-q+u]=\Big(\mathcal{R}_{L}\Big(\int_{0}^{q}\rmd u\,h(u)f(\cdot+\tfrac{u}{L})\Big)\Big)[\mu,d-q]+L\int_{0}^{q}\rmd u\,h(u)\int_{\mu}^{\mu+\frac{u}{L}}\rmd v\,f(v)\;.
\end{equation}
Applying this property to $h(u)=f(u)=u^{-1/2}$ and using (\ref{d log(sqrt-sqrt)}) to integrate inside the operator $\mathcal{R}_{L}$, one has
\begin{eqnarray}
&& \pm\frac{1}{2\sqrt{L}}\int_{0}^{q}\frac{\rmd u}{\sqrt{u}}\Big(\mathcal{R}_{L}\frac{1}{\sqrt{\cdot}}\Big)[\mu,d-q+u]
+\frac{(\mathcal{R}_{L}\log)[\mu,d-q]}{2}\\
&& =\Big(\mathcal{R}_{L}\log\big(\sqrt{\cdot}\pm\sqrt{q/L}\big)\Big)[\mu,d]
-L\int_{0}^{q/L}\rmd u\,\log(\sqrt{\mu}\pm\sqrt{u})\;.\nonumber
\end{eqnarray}
After some simplifications, we arrive at (\ref{EM sqrt(Stirling)}).
\end{subsubsection}

\begin{subsubsection}{Singularities at both ends}
In all the cases described so far in this section, we observe that the asymptotic expansion can always be written as
\begin{equation}
\label{EM regular singular}
\sum_{j=1}^{M}f\Big(\frac{j+d}{L}\Big)
\simeq L\Big(\int_{0}^{\mu}\rmd u\,f(u)\Big)
+(\mathcal{R}_{L}f)[\mu,d]
+(\mathcal{S}_{L}f)[d]\;,
\end{equation}
with some regularization needed when the integral does not converge. This is true in general since one can always decompose the sum from $1$ to $M$ as a sum from $1$ to $\varepsilon L$ plus a sum from $\varepsilon L+1$ to $M$. For all $\varepsilon>0$ such that $\varepsilon L$ is an integer, the asymptotics of the second sum is given by (\ref{EM regular}) and the limit $\varepsilon\to0$ can be written as (\ref{EM regular singular}) with a singular part $(\mathcal{S}_{L}f)[d]$ independent of $\mu$.

Let us now consider the case of a function $f$, with singularities at both $0$ and $\rho=N/L$. The singularities are specified by functions $\mathcal{S}_{0}=\mathcal{S}_{L}f$ and $\mathcal{S}_{\rho}=\mathcal{S}_{L}f(\rho-\cdot)$ in (\ref{EM regular singular}). Splitting the sum into two parts at $M=\mu L$ leads to
\begin{equation}
\sum_{j=1}^{N}f\Big(\frac{j+d}{L}\Big)=\sum_{j=1}^{M}f\Big(\frac{j+d}{L}\Big)+\sum_{j=1}^{N-M}f\Big(\rho-\frac{j-d-1}{L}\Big)\;.
\end{equation}
One can use (\ref{EM regular singular}) on both parts. From (\ref{Bernoulli symmetry}), the regular remainder terms at $\mu$ cancels: $(\mathcal{R}_{L}f)[\mu,d]
+(\mathcal{R}_{L}f(\rho-\cdot))[\rho-\mu,-d-1]=0$, leaving only the integral and the singular terms:
\begin{equation}
\sum_{j=1}^{N}f\Big(\frac{j+d}{L}\Big)
\simeq L\Big(\int_{0}^{\rho}\rmd u\,f(u)\Big)+\mathcal{S}_{0}[d]+\mathcal{S}_{\rho}[-d-1]\;.
\end{equation}
In particular, for
\begin{equation}
\label{f[fk]}
f(x)=\alpha\log x+\sum_{k=-1}^{\infty}f_{k}x^{k/2}=\overline{\alpha}\log(\rho-x)+\sum_{k=-1}^{\infty}\overline{f}_{k}(\rho-x)^{k/2}\;,
\end{equation}
one has the asymptotic expansion
\begin{eqnarray}
\label{EM sqrt sqrt}
&& \sum_{j=1}^{N}f\Big(\frac{j+d}{L}\Big)\simeq L\Big(\int_{0}^{\rho}\rmd u\,f(u)\Big)
+(\alpha-\overline{\alpha})(d+\half)\log L
+(\alpha+\overline{\alpha})\log\sqrt{2\pi}
-\alpha\log\Gamma(d+1)\nonumber\\
&&\hspace{25mm} -\overline{\alpha}\log\Gamma(-d)+\sum_{k=-1}^{\infty}f_{k}\,\frac{\zeta(-k/2,d+1)}{L^{k/2}}+\sum_{k=-1}^{\infty}\overline{f}_{k}\,\frac{\zeta(-k/2,-d)}{L^{k/2}}\;.
\end{eqnarray}

Similarly let us consider a function $f$ with square root singularities and singularities as a sum of two square roots, both at $0$ and $\rho$:
\begin{equation}
f(x)=\Big(\sqrt{x}+\sigma_{0}\sqrt{\frac{q_{0}}{L}}\Big)h_{0}(x)=\Big(\sqrt{\rho-x}+\sigma_{1}\sqrt{\frac{q_{1}}{L}}\Big)h_{1}(\rho-x)\;.
\end{equation}
The parameters $q_{0}$ and $q_{1}$ are complex numbers, $\sigma_{0}$ and $\sigma_{1}$ are equal to $1$ or $-1$. The functions $h_{0}$ and $h_{1}$ have only square root singularities at $0$:
\begin{equation}
h_{0}(x)=\exp\Big(\sum_{r=0}^{\infty}h_{0,r}x^{r/2}\Big)
\quad\text{and}\quad
h_{1}(x)=\exp\Big(\sum_{r=0}^{\infty}h_{1,r}x^{r/2}\Big)\;,
\end{equation}
with coefficients $h_{0,r}$ and $h_{1,r}$ which may depend on $L$. Using (\ref{EM sqrt(Stirling)}) and (\ref{EM power}), one finds after some simplifications the asymptotic expansion
\begin{eqnarray}
\label{EM sqrt(Stirling) sqrt(Stirling)}
&& \sum_{j=m_{0}}^{N-m_{1}}\log f\Big(\frac{j+d}{L}\Big)
\simeq L\Big(\int_{0}^{\rho}\rmd u\,\log f(u)\Big)
+\frac{m_{0}+m_{1}-1}{2}\,\log L+\log\sqrt{2\pi}\nonumber\\
&&\hspace{1mm} +\frac{q_{0}}{2}-\frac{q_{0}\log q_{0}}{2}+\frac{\rmi\pi(1-\sigma_{0})q_{0}}{2}\,\sgn(\arg q_{0})-\frac{\log\Gamma(m_{0}+d-q_{0})}{2}+\sigma_{0}\int_{0}^{q_{0}}\rmd u\,\frac{\zeta(\frac{1}{2},m_{0}+d-q_{0}+u)}{2\sqrt{u}}\nonumber\\
&&\hspace{1mm} +\frac{q_{1}}{2}-\frac{q_{1}\log q_{1}}{2}+\frac{\rmi\pi(1-\sigma_{1})q_{1}}{2}\,\sgn(\arg q_{1})-\frac{\log\Gamma(m_{1}-d-q_{1})}{2}+\sigma_{1}\int_{0}^{q_{1}}\rmd u\,\frac{\zeta(\frac{1}{2},m_{1}-d-q_{1}+u)}{2\sqrt{u}}\nonumber\\
&&\hspace{1mm} +\sum_{r=0}^{\infty}\frac{h_{0,r}\,\zeta(-r/2,m_{0}+d)}{L^{r/2}}
+\sum_{r=0}^{\infty}\frac{h_{1,r}\,\zeta(-r/2,m_{1}-d)}{L^{r/2}}\;.
\end{eqnarray}
The integers $m_{0}$ and $m_{1}$ were added in order to treat singularities that may appear for $j$ close to $1$ and $N$ such that $f((j+d)/L)=0$. Their contribution to (\ref{EM sqrt(Stirling) sqrt(Stirling)}) come from (\ref{sum_power[zeta]}) and (\ref{sum_log_sqrt[Gamma,zeta]}).

We assumed that for large $L$, the coefficients $h_{0,r}$ and $h_{1,r}$ do not grow too fast when $r$ increases. In this paper, we only use (\ref{EM sqrt(Stirling) sqrt(Stirling)}) with coefficients $h_{0,r}$ and $h_{1,r}$ that have a finite limit when $L$ goes to $\infty$, with an expansion in powers of $1/\sqrt{L}$. Also, we only need the expansion up to order $L^{0}$. One has
\begin{eqnarray}
&& \sum_{r=0}^{\infty}\frac{h_{0,r}\,\zeta(-r/2,m_{0}+d)}{L^{r/2}}
+\sum_{r=0}^{\infty}\frac{h_{1,r}\,\zeta(-r/2,m_{1}-d)}{L^{r/2}}\\
&& =\frac{1-m_{0}-m_{1}}{2}\,\log L
+(\half-m_{0}-d)\log\frac{\sigma_{0}f(0)}{\sqrt{q_{0}}}
+(\half-m_{1}+d)\log\frac{\sigma_{1}f(\rho)}{\sqrt{q_{1}}}
+\O\Big(\frac{1}{\sqrt{L}}\Big)\;.\nonumber
\end{eqnarray}
\end{subsubsection}
\end{subsection}

\begin{subsection}{Two-dimensional sums}
Generalizations of the Euler-Maclaurin formula can also be used in the case of summations over two indices. Things are however more complicated than in the one-dimensional case because the way to handle those sums depends on the two-dimensional domain of summation, and because of the new kinds of singularities that can happen at singular points of the boundary. We consider here only the case of rectangles $\{(j,j'),1\leq j\leq M,1\leq j'\leq M'\}$ and triangles $\{(j,j'),1\leq j<j'\leq M\}$ that are needed for the asymptotic expansion of the normalization.

\begin{subsubsection}{Rectangle with square root singularities at a corner}
We consider a function of two variables $f$ with square root singularities at the point $(0,0)$
\begin{equation}
\label{f sqrt sqrt}
f(u,v)=\sum_{k=0}^{\infty}\sum_{k'=0}^{\infty}f_{k,k'}u^{k/2}v^{k'/2}\;,
\end{equation}
and define two auxiliary functions on the edges of the square
\begin{equation}
g_{k}(v)=\sum_{k'=0}^{\infty}f_{k,k'}v^{k'/2}
\quad\text{and}\quad
h_{k'}(u)=\sum_{k=0}^{\infty}f_{k,k'}u^{k/2}\;.
\end{equation}
Then, taking $M=\mu L$ and $N=\rho L$, (\ref{sum_power[zeta]}) gives the large $L$ asymptotic expansion
\begin{eqnarray}
\label{EM rectangle sqrt}
&& \sum_{j=1}^{M}\sum_{j'=1}^{N}f\Big(\frac{j+d}{L},\frac{j'+d'}{L}\Big)
\simeq L^{2}\int_{0}^{\mu}\rmd u\int_{0}^{\rho}\rmd v\,f(u,v)\nonumber\\
&& +\sum_{\ell=1}^{\infty}\frac{B_{\ell}(d+1)}{\ell!L^{\ell-2}}\int_{0}^{\rho}\rmd v\,f^{(\ell-1,0)}(\mu,v)
+\sum_{\ell=1}^{\infty}\frac{B_{\ell}(d'+1)}{\ell!L^{\ell-2}}\int_{0}^{\mu}\rmd u\,f^{(0,\ell-1)}(u,\rho)\nonumber\\
&& +\sum_{k=0}^{\infty}\frac{\zeta(-k/2,d+1)}{L^{\frac{k}{2}-1}}\int_{0}^{\rho}\rmd v\,g_{k}(v)
+\sum_{k=0}^{\infty}\frac{\zeta(-k/2,d'+1)}{L^{\frac{k}{2}-1}}\int_{0}^{\mu}\rmd u\,h_{k}(u)\\
&& +\sum_{\ell,\ell'=1}^{\infty}\frac{B_{\ell}(d+1)}{\ell!L^{\ell-1}}\,\frac{B_{\ell'}(d'+1)}{\ell'!L^{\ell'-1}}\,f^{(\ell-1,\ell'-1)}(\mu,\rho)
+\sum_{k=0}^{\infty}\frac{\zeta(-k/2,d+1)}{L^{\frac{k}{2}}}\sum_{\ell=1}^{\infty}\frac{B_{\ell}(d'+1)}{\ell!L^{\ell-1}}\,g_{k}^{(\ell-1)}(\rho)\nonumber\\
&& +\sum_{k=0}^{\infty}\frac{\zeta(-k/2,d'+1)}{L^{\frac{k}{2}}}\sum_{\ell=1}^{\infty}\frac{B_{\ell}(d+1)}{\ell!L^{\ell-1}}\,h_{k}^{(\ell-1)}(\mu)
+\sum_{k,k'=0}^{\infty}\frac{\zeta(-k/2,d+1)\zeta(-k/2,d'+1)}{L^{\frac{k+k'}{2}}}\,f_{k,k'}\;.\nonumber
\end{eqnarray}
The first term with the double integral is related to the full square, the next four terms with a single integral to the four edges of the square, and the four last terms to the four corners of the square.
\end{subsubsection}

\begin{subsubsection}{Triangle with square root singularities at a corner}
We consider again a function of two variables $f$ with square root singularities at $(0,0)$ as in (\ref{f sqrt sqrt}), and define
\begin{equation}
f_{k}(v)=\sum_{k'=0}^{\infty}f_{k,k'}v^{k'/2}\;.
\end{equation}
One has the asymptotic expansion
\begin{eqnarray}
\label{EM triangle sqrt}
&& \sum_{j=1}^{M}\sum_{j'=j+1}^{M}f\Big(\frac{j+d}{L},\frac{j'+d'}{L}\Big)
\simeq L^{2}\int_{0}^{\mu}\rmd u\int_{u}^{\mu}\rmd v\,f(u,v)\\
&& +\sum_{\ell=1}^{\infty}\frac{B_{\ell}(d'+1)}{\ell!L^{\ell-2}}\,\partial_{\mu}^{\ell-1}\int_{0}^{\mu}\rmd u\,f(u,\mu)
+\sum_{k=0}^{\infty}\frac{\zeta(-k/2,d+1)}{L^{\frac{k}{2}-1}}\int_{0}^{\mu}\rmd v\,f_{k}(v)\nonumber\\
&&\hspace{55mm} +\sum_{\ell=1}^{\infty}\frac{B_{\ell}(d-d')}{\ell!L^{\ell-2}}\,\mint_{0}^{\mu}\rmd u\,f^{(\ell-1,0)}(u,u)\nonumber\\
&& +\sum_{\ell,\ell'=1}^{\infty}\frac{B_{\ell}(d-d')}{\ell!L^{\ell-1}}\,\frac{B_{\ell'}(d'+1)}{\ell'!L^{\ell'-1}}\,\partial_{\mu}^{\ell'-1}f^{(\ell-1,0)}(\mu,\mu)
+\sum_{k=0}^{\infty}\frac{\zeta(-k/2,d+1)}{L^{\frac{k}{2}}}\sum_{\ell=1}^{\infty}\frac{B_{\ell}(d'+1)}{\ell!L^{\ell-1}}\,f_{k}^{(\ell-1)}(\mu)\nonumber\\
&&\hspace{75mm} +\sum_{k,k'=0}^{\infty}\frac{f_{k,k'}}{L^{\frac{k+k'}{2}}}\,\tilde{\zeta}_{0}(-k/2,-k'/2;d+1,d'+1)\;.\nonumber
\end{eqnarray}
The modified integral is defined as in (\ref{mint}), after expanding near $u=0$. The modified double Hurwitz zeta function $\tilde{\zeta}_{0}$ is defined in appendix \ref{appendix zeta}. The first term in (\ref{EM triangle sqrt}) corresponds to the whole triangle, the next three terms to the three edges, and the last three terms to the three corners.

As usual, (\ref{EM triangle sqrt}) can be shown by expanding $f$ near the point $(0,0)$. At each order in the expansion, the summation over $j,j'$ can be performed in terms of double Hurwitz zeta functions using
\begin{eqnarray}
\label{sum_power[zeta double]}
&& \sum_{j=1}^{M}\sum_{j'=j+1}^{M}(j+d)^{\nu}(j'+d')^{\nu'}=\zeta(-\nu,-\nu';d+1,d'+1)\\
&&\hspace{5mm} +\zeta(-\nu',-\nu;M+d'+1,M+d)-\zeta(-\nu,d+1)\zeta(-\nu',M+d'+1)\;.\nonumber
\end{eqnarray}
The summation formula (\ref{sum_power[zeta double]}) can be shown by using the decomposition
\begin{equation}
  \begin{picture}(115,25)
    \put(0,0){\color{lightgray}\polygon*(0,0)(10,10)(0,10)}
    \put(0,0){\polygon(0,0)(10,10)(0,10)}
    \put(30,0){\color{lightgray}\polygon*(0,0)(25,25)(0,25)}
    \put(30,0){\polygon(0,0)(10,10)(0,10)}
    \put(55,0){\color{lightgray}\polygon*(10,10)(25,25)(25,10)}
    \put(55,0){\polygon(0,0)(10,10)(0,10)}
    \put(90,0){\color{lightgray}\polygon*(0,10)(25,10)(25,25)(0,25)}
    \put(90,0){\polygon(0,0)(10,10)(0,10)}
    \put(20,3){$=$}
    \put(45,3){$+$}
    \put(80,3){$-$}
  \end{picture}\;,
\end{equation}
writing
\begin{eqnarray}
&& \sum_{j=1}^{M}\sum_{j'=j+1}^{M}(j+d)^{\nu}(j'+d')^{\nu'}
=\sum_{j=1}^{\infty}\sum_{j'=j+1}^{\infty}(j+d)^{\nu}(j'+d')^{\nu'}\\
&& +\sum_{j'=M+1}^{\infty}\sum_{j=j'}^{\infty}(j+d)^{\nu}(j'+d')^{\nu'}
-\sum_{j=1}^{\infty}\sum_{j'=M+1}^{\infty}(j+d)^{\nu}(j'+d')^{\nu'}\;,\nonumber
\end{eqnarray}
provided that $\nu'<-1$ and $\nu+\nu'<-2$ to ensure the convergence of the infinite sums. From the definition (\ref{Hurwitz zeta double}) of double Hurwitz zeta functions, this leads to (\ref{sum_power[zeta double]}). By analytic continuation, (\ref{sum_power[zeta double]}) is valid for all $\nu$, $\nu'$ different from the poles of simple and double $\zeta$. It is also valid when replacing the double $\zeta$ by their modified values $\tilde{\zeta}_{\alpha}$ and $\tilde{\zeta}_{1-\alpha}$ when $2-s-s'\in\mathbb{N}$. Indeed, using (\ref{Bernoulli symmetry}), we observe that the quantity $\tilde{\zeta}_{\alpha}(s,s';z,z')+\tilde{\zeta}_{1-\alpha}(s',s;M+z',M+z-1)$ is not divergent on the line $2-s-s'\in\mathbb{N}$, and is independent of $\alpha$. One has
\begin{equation}
\lim_{s+s'\to2-n}(\zeta(s,s';z,z')-\zeta(s',s;M+z',M+z-1))
=\tilde{\zeta}_{\alpha}(s,s';z,z')-\tilde{\zeta}_{1-\alpha}(s',s;M+z',M+z-1)\;
\end{equation}
for arbitrary direction in the convergence of $s+s'$ to $2-n$ and for arbitrary $\alpha$.

Expanding for large $M=\mu L$ using (\ref{Hurwitz zeta asymptotics}) and (\ref{Hurwitz zeta double asymptotics}), a tedious calculation finally leads to the large $L$ asymptotic expansion (\ref{EM triangle sqrt}).
\end{subsubsection}

\begin{subsubsection}{Triangle with square root singularities at all corners}
We consider a function $f(u,v)$ of two variables, analytic in the interior of the domain $\{(u,v),0<u<v<\rho\}$ and with square root singularities at the points $(0,0)$, $(\rho,\rho)$, $(0,\rho)$. We define $g(u,v)=f(\rho-v,\rho-u)$ and $h(u,v)=f(u,\rho-v)$. The expansions near the singularities are given by
\begin{equation}
f(u,v)=\sum_{k=0}^{\infty}\sum_{k'=0}^{\infty}f_{k,k'}u^{k/2}v^{k'/2}\;,\quad
g(u,v)=\sum_{k=0}^{\infty}\sum_{k'=0}^{\infty}g_{k,k'}u^{k/2}v^{k'/2}\;,\quad
h(u,v)=\sum_{k=0}^{\infty}\sum_{k'=0}^{\infty}h_{k,k'}u^{k/2}v^{k'/2}\;.
\end{equation}
We also define
\begin{equation}
f_{\nu}(v)=\sum_{k'=0}^{\infty}f_{\nu,k'}v^{k'/2}\;,\qquad
g_{\nu}(v)=\sum_{k'=0}^{\infty}g_{\nu,k'}v^{k'/2}\;.
\end{equation}
In order to handle the singularities at the corners, we decompose the triangle as
\begin{eqnarray}
&& \sum_{j=1}^{N}\sum_{j'=j+1}^{N}f\Big(\frac{j+d}{L},\frac{j'+d'}{L}\Big)
=\sum_{j=1}^{M}\sum_{j'=j+1}^{M}f\Big(\frac{j+d}{L},\frac{j'+d'}{L}\Big)\\
&& +\sum_{j'=1}^{N-M}\sum_{j=j'+1}^{N-M}g\Big(\frac{j'-d'-1}{L},\frac{j-d-1}{L}\Big)
+\sum_{j=1}^{M}\sum_{j'=1}^{N-M}h\Big(\frac{j+d}{L},\frac{j'-d'-1}{L}\Big)\;.\nonumber
\end{eqnarray}
Then, for large $N=\rho L$ and $M=\mu L$, using (\ref{EM rectangle sqrt}) and (\ref{EM triangle sqrt}) and combining all the terms leads to the large $L$ asymptotic expansion
\begin{eqnarray}
\label{EM triangle sqrt sqrt sqrt}
&& \sum_{j=1}^{N}\sum_{j'=j+1}^{N}f\Big(\frac{j+d}{L},\frac{j'+d'}{L}\Big)\simeq L^{2}\int_{0}^{\rho}\rmd u\int_{u}^{\rho}\rmd v\,f(u,v)\\
&& +\sum_{k=0}^{\infty}\frac{\zeta(-k/2,d+1)}{L^{k/2-1}}\int_{0}^{\rho}\rmd v\,f_{k}(v)
+\sum_{k=0}^{\infty}\frac{\zeta(-k/2,-d')}{L^{k/2-1}}\int_{0}^{\rho}\rmd v\,g_{k}(v)\nonumber\\
&& +\sum_{\ell=1}^{\infty}\frac{B_{\ell}(d-d')}{\ell!L^{\ell-2}}\Big(\mint_{0}^{\mu}\rmd v\,f^{(l-1,0)}(v,v)+\sum_{m=1}^{\ell-1}(-1)^{\ell-m}f^{(m-1,l-m-1)}(\mu,\mu)\nonumber\\
&&\hspace{70mm} +(-1)^{\ell-1}\mint_{\mu}^{\rho}\rmd v\,f^{(0,l-1)}(v,v)\Big)\nonumber\\
&& +\sum_{k=0}^{\infty}\sum_{k'=0}^{\infty}\frac{f_{k,k'}}{L^{\frac{k+k'}{2}}}\,\tilde{\zeta}_{0}(-k/2,-k'/2,d+1,d'+1)
+\sum_{k=0}^{\infty}\sum_{k'=0}^{\infty}\frac{g_{k',k}}{L^{\frac{k+k'}{2}}}\,\tilde{\zeta}_{0}(-k'/2,-k/2,-d',-d)\nonumber\\
&&\hspace{72mm} +\sum_{k=0}^{\infty}\sum_{k'=0}^{\infty}\frac{h_{k,k'}}{L^{\frac{k+k'}{2}}}\,\zeta(-k/2,d+1)\zeta(-k'/2,-d')\;.\nonumber
\end{eqnarray}
The modified integral is defined as in (\ref{mint}), after expanding near $v=0$ and $v=\rho$. The expansion is independent of the arbitrary parameter $\mu$, $0<\mu<\rho$ that splits the modified integral.
\end{subsubsection}

\begin{subsubsection}{Triangle with logarithmic singularity on an edge: Barnes function}
A two dimensional generalization of Stirling's formula for the $\Gamma$ function is
\begin{eqnarray}
\label{EM triangle log}
&& \sum_{j=1}^{N}\sum_{j'=j+1}^{N}\log(j'-j+d)\simeq\frac{N^{2}\log N}{2}-\frac{3N^{2}}{4}+d\,N\log N\\
&& +N\big(\log\sqrt{2\pi}-d-\log\Gamma(d+1)\big)+\Big(\frac{d^{2}}{2}-\frac{1}{12}\Big)\log N+\zeta'(-1)+d\log\sqrt{2\pi}-\log G(d+1)\;,\nonumber
\end{eqnarray}
where $\zeta$ is Riemann's zeta function and $\log G$ is the log Barnes function, equal to the analytic continuation of the logarithm of the Barnes function $G$. The constant term is usually written in terms of the Glaisher-Kinkelin constant $A=\exp(\tfrac{1}{12}-\zeta'(-1))$. The expansion (\ref{EM triangle log}) follows from the identity $G(z+1)=\Gamma(z)G(z)$ and the asymptotics of Barnes function for large argument
\begin{equation}
\label{Barnes G asymptotics}
\log G(N+1)\simeq\frac{N^{2}\log N}{2}-\frac{3N^{2}}{4}+\frac{\log(2\pi)}{2}\,N-\frac{\log N}{12}+\zeta'(-1)\;.
\end{equation}
\end{subsubsection}

\begin{subsubsection}{Square with logarithm of a sum of square roots}
We consider two dimensional generalizations of the $\sqrt{\text{Stirling}}$ formula (\ref{EM sqrt(Stirling)}). One has the asymptotics
\begin{eqnarray}
\label{EM rectangle log(sqrt+sqrt)}
&& \sum_{j=1}^{N}\sum_{j'=1}^{N}\log(\sqrt{j+d}+\sqrt{j'+d})\simeq
\frac{N^{2}\log N}{2}+\frac{N^{2}}{4}+(2d+1)N\\
&&\hspace{7mm} +4\sqrt{N}\zeta(-\half,d+1)-\frac{\log N}{24}
+(d+\half)^{2}+\int_{0}^{d+\half}\rmd u\,\frac{\zeta(\half,\half+u)^{2}}{2}+\kappa_{0}\;,\nonumber
\end{eqnarray}
with $\kappa_{0}\approx-0.128121307412384$. In order to show this, we introduce $M=\mu N$, $0<\mu<1$ and decompose the sum as
\begin{equation}
\sum_{j=1}^{N}\sum_{j'=1}^{N}
\;=\;\sum_{j=1}^{M}\sum_{j'=1}^{M}
\;+\sum_{j=M+1}^{N}\sum_{j'=1}^{M}
\;+\;\sum_{j=1}^{M}\sum_{j'=M+1}^{N}
+\sum_{j=M+1}^{N}\sum_{j'=M+1}^{N}\;.
\end{equation}
The last three terms in the right hand side can be evaluated using Euler-Maclaurin formula in a rectangle with only square root singularities, using (\ref{EM rectangle sqrt}) and
\begin{eqnarray}
&& \log(\sqrt{u+\alpha}+\sqrt{v+\beta})\\
&& =\partial_{u}\Big(-\frac{u}{2}+\sqrt{u+\alpha}\sqrt{v+\beta}+(u+\alpha-v-\beta)\log(\sqrt{u+\alpha}+\sqrt{v+\beta})\Big)\nonumber\\
&& =\partial_{u}\partial_{v}\Big(-\frac{3uv}{4}+\frac{(u+\alpha)^{3/2}\sqrt{v+\beta}}{2}+\frac{\sqrt{u+\alpha}(v+\beta)^{3/2}}{2}-\frac{(u+\alpha-v-\beta)^{2}}{2}\log(\sqrt{u+\alpha}+\sqrt{v+\beta})\Big)\;\nonumber
\end{eqnarray}
to compute the integrals. One finds up to order $0$ in $N$
\begin{eqnarray}
&& \sum_{j=1}^{N}\sum_{j'=1}^{N}\log(\sqrt{j+d}+\sqrt{j'+d'})
-\sum_{j=1}^{M}\sum_{j'=1}^{M}\log(\sqrt{j+d}+\sqrt{j'+d'})\\
&& \simeq\Big(\frac{1}{4}-\frac{\mu^{2}}{4}-\frac{\mu^{2}\log\mu}{2}\Big)N^{2}
+(2d+1)(1-\mu)N+4(1-\sqrt{\mu})\zeta(-\half,d+1)\sqrt{N}+\frac{\log\mu}{24}\;.\nonumber
\end{eqnarray}
The limit $\mu\to0$ leads to the divergent terms in (\ref{EM rectangle log(sqrt+sqrt)}). The term at order $N^{0}$ follows from the summation formula (\ref{sum_power[zeta]}) applied to the identity
\begin{equation}
\partial_{\lambda}\log\big(\sqrt{j+d+\lambda}+\sqrt{j'+d+\lambda}\big)=\frac{1}{2\sqrt{j+d+\lambda}\sqrt{j'+d+\lambda}}\;.
\end{equation}
The remaining constant of integration $\kappa_{0}$ can be evaluated numerically with high precision using BST algorithm, as described in section \ref{section numerics}.
\end{subsubsection}

\begin{subsubsection}{Square with logarithm of a sum of square roots (2)}
One has the asymptotics
\begin{eqnarray}
\label{EM rectangle log(sqrt+sqrt) i}
&& \sum_{j=1}^{N}\sum_{j'=1}^{N}\log\big(\sqrt{-\rmi(j+d)}+\sqrt{\rmi(j'+d')}\big)
\simeq\frac{N^{2}\log N}{2}+\Big(-\frac{3}{4}+\log2\Big)N^{2}\\
&& +\big((d+d'+1)\log2-\rmi(d-d')\big)N-2\rmi\sqrt{N}\big(\zeta(-\half,d+1)-\zeta(-\half,d'+1)\big)+\Big(\frac{1}{24}-\frac{(d+d'+1)^{2}}{4}\Big)\log N\nonumber\\
&& -\frac{\rmi(d-d')(d+d'+1)}{2}+\int_{d'+\half}^{0}\rmd u\,\frac{\zeta(\half,d+1+u)\zeta(\half,d'+1-u)}{2\rmi}+\kappa_{1}(d+d')\;.\nonumber
\end{eqnarray}
When $d+d'=-1$, the constant of integration is $\kappa_{1}(-1)\approx0.05382943932689441$. The derivation is essentially identical to the one of (\ref{EM rectangle log(sqrt+sqrt)}).
\end{subsubsection}
\end{subsection}
\end{section}

\begin{section}{Large \texorpdfstring{$L$}{L} asymptotics}
\label{section asymptotics}
In this section, we compute the large $L$ asymptotics of the quantities
\begin{eqnarray}
\label{Xi1}
&& \Xi_{1}=\frac{1}{N}\sum_{j=1}^{N}\frac{y_{j}}{\rho+(1-\rho)y_{j}}\;,\\
\label{Xi2}
&& \Xi_{2}=\prod_{j=1}^{N}\Big(1+\frac{1-\rho}{\rho}\,y_{j}\Big)\;,\\
\label{Xi3}
&& \Xi_{3}=\prod_{j=1}^{N}\prod_{k=j+1}^{N}\frac{y_{j}-y_{k}}{y_{j}^{0}-y_{k}^{0}}\;,\\
\label{Xi4}
&& \Xi_{4}=\prod_{j=1}^{N}\prod_{k=j+1}^{N}\big(y_{j}^{0}-y_{k}^{0}\big)\;,
\end{eqnarray}
for $y_{j}$'s given by (\ref{y[k]}), $k_{j}$'s constructed in terms of sets $A_{0}^{\pm}$, $A^{\pm}$, and with the correction (\ref{b[c]}) to $b$. The parameter $c$ will be in this section an arbitrary complex number that is \textbf{not} required to verify (\ref{phi(c)}).

\begin{subsection}{Function \texorpdfstring{$\Phi$}{Phi}}
We introduce the function $\Phi$ defined from $g$ (\ref{g}) by
\begin{equation}
\label{Phi}
\Phi(u)=g^{-1}\big(\rme^{-b_{0}+2\rmi\pi u}\big)\;.
\end{equation}
Writing the parameter $b$ as in (\ref{b[c]}), the Bethe roots can be expressed as
\begin{equation}
y_{j}=\Phi\Big(\frac{k_{j}+\rmi c}{L}\Big)\;.
\end{equation}
In particular, for the stationary eigenstate, one has
\begin{equation}
y_{j}^{0}=\Phi\Big(-\frac{\rho}{2}+\frac{j-\frac{1}{2}+\rmi c}{L}\Big)\;.
\end{equation}
For the first eigenstates, the $k_{j}$'s added ($\pm(N/2+a)$, $a\in A^{\pm}$) correspond to $y_{j}=\Phi(\pm\tfrac{\rho}{2}+\frac{\rmi(c\mp\rmi a)}{L})$, while the $k_{j}$'s removed ($\pm(N/2-a)$, $a\in A_{0}^{\pm}$) correspond to $y_{j}=\Phi(\pm\tfrac{\rho}{2}+\frac{\rmi(c\pm\rmi a)}{L})$. These expressions are suited for the use of the Euler-Maclaurin formula to compute the asymptotics of sums of the form $\sum_{j=1}^{N}f(y_{j})$ and $\sum_{j=1}^{N}\sum_{j'=j+1}^{N}f(y_{j},y_{j'})$.

The function $\Phi$ verifies $\Phi(\pm\rho/2)=-\tfrac{\rho}{1-\rho}$. The points $\pm\rho/2$ are branch points of the function $\Phi$. The expansion of $\Phi$ around them is given by
\begin{eqnarray}
\label{Phi expansion}
&& \Phi\big(\pm(\tfrac{\rho}{2}-u)\big)\simeq
-\frac{\rho}{1-\rho}\bigg(
1-\frac{\sqrt{2}(1\mp\rmi)\sqrt{\pi}\sqrt{u}}{\sqrt{\rho(1-\rho)}}
\mp\frac{4\rmi\pi(1+\rho)u}{3\rho(1-\rho)}\\
&&\hspace{41mm} +\frac{\sqrt{2}(1\pm\rmi)\pi^{3/2}(1+11\rho+\rho^{2})u^{3/2}}{9(\rho(1-\rho))^{3/2}}
+\frac{8\pi^{2}(1+\rho)(1-25\rho+\rho^{2})u^{2}}{135\rho^{2}(1-\rho)^{2}}
\bigg)\;.\nonumber
\end{eqnarray}

A useful property of the function $\Phi$ is that its derivative can be expressed in terms of $\Phi$ alone. One has
\begin{equation}
\label{Phi'}
\Phi'(u)=-2\rmi\pi\,\frac{\Phi(u)(1-\Phi(u))}{\rho+(1-\rho)\Phi(u)}\;.
\end{equation}
\end{subsection}

\begin{subsection}{Asymptotics of \texorpdfstring{$\Xi_{1}$}{Xi1}}
The Bethe roots $y_{j}$ can be replaced by $y_{j}^{0}$ in (\ref{Xi1}), up to corrections obtained by summing over the sets $A_{0}^{\pm}$, $A^{\pm}$. Writing $b$ as (\ref{b[c]}), the summand is equal to $y_{j}/(\rho+(1-\rho)y_{j})=f(\rho/2+(k_{j}+\rmi c)/L)$ with $f(u)=\Phi(u-\tfrac{\rho}{2})/(\rho+(1-\rho)\Phi(u-\tfrac{\rho}{2}))$. One has
\begin{eqnarray}
\label{Xi1[A]}
&& \sum_{j=1}^{N}\frac{y_{j}}{\rho+(1-\rho)y_{j}}
=\sum_{j=1}^{N}f\Big(\frac{j-\frac{1}{2}+\rmi c}{L}\Big)
+\sum_{a\in A^{-}}f\Big(\frac{\rmi(c+\rmi a)}{L}\Big)
-\sum_{a\in A_{0}^{-}}f\Big(\frac{\rmi(c-\rmi a)}{L}\Big)\\
&&\hspace{56mm} +\sum_{a\in A^{+}}f\Big(\rho+\frac{\rmi(c-\rmi a)}{L}\Big)
-\sum_{a\in A_{0}^{+}}f\Big(\rho+\frac{\rmi(c+\rmi a)}{L}\Big)\;.\nonumber
\end{eqnarray}
The function $f$ verifies (\ref{f[fk]}) with first coefficients equal to $\alpha=\overline{\alpha}=0$ and
\begin{equation}
f_{-1}=-\frac{(1-\rmi)\sqrt{\rho}}{2^{3/2}\sqrt{\pi}\sqrt{1-\rho}}
\quad\text{and}\quad
\overline{f}_{-1}=-\frac{(1+\rmi)\sqrt{\rho}}{2^{3/2}\sqrt{\pi}\sqrt{1-\rho}}\;.
\end{equation}
From (\ref{EM sqrt sqrt}), the expansion up to order $\sqrt{L}$ of the sum over $j$ in the right hand side of (\ref{Xi1[A]}) is
\begin{equation}
\sum_{j=1}^{N}f\Big(\frac{j-\frac{1}{2}+\rmi c}{L}\Big)
\simeq L\Big(\int_{0}^{\rho}\rmd u\,f(u)\Big)
+\sqrt{L}\big(f_{-1}\zeta(\tfrac{1}{2},\tfrac{1}{2}+\rmi c)+\overline{f}_{-1}\zeta(\tfrac{1}{2},\tfrac{1}{2}-\rmi c)\big)\;.
\end{equation}
The integral can be computing by making the change of variables $z=\Phi(u-\tfrac{\rho}{2})$, as explained in appendix \ref{appendix integrals}. The residue calculation gives $\int_{0}^{\rho}\rmd u\,f(u)=0$.

At leading order in $L$, using (\ref{Phi expansion}) to treat the sums over $a$, and (\ref{zeta'}), we can express $\Xi_{1}$ at leading order in terms of the derivative of the function $\varphi_{r}$ (\ref{phi[A,zeta]}). We find
\begin{equation}
\label{Asymptotics 1}
\frac{1}{N}\sum_{j=1}^{N}\frac{y_{j}}{\rho+(1-\rho)y_{j}}\simeq\frac{\varphi_{r}'(2\pi c)}{\sqrt{\rho(1-\rho)}\,\sqrt{L}}\;.
\end{equation}
\end{subsection}

\begin{subsection}{Asymptotics of \texorpdfstring{$\Xi_{2}$}{Xi2}}
We consider the logarithm of $\Xi_{2}$, defined in (\ref{Xi2}). The calculation of the large $L$ asymptotics follows closely the one for $\Xi_{1}$, except one needs to push the expansion of the sum up to the constant term in $L$ in order to get the prefactor of the exponential in $\Xi_{2}$. The summand is equal to $\log(1+y_{j}(1-\rho)/\rho)=f(\rho/2+(k_{j}+\rmi c)/L)$ with $f(u)=\log(1+\Phi(u-\tfrac{\rho}{2})(1-\rho)/\rho)$. One has again
\begin{eqnarray}
\label{Xi2[A]}
&& \sum_{j=1}^{N}\log\Big(1+\frac{1-\rho}{\rho}\,y_{j}\Big)
=\sum_{j=1}^{N}f\Big(\frac{j-\frac{1}{2}+\rmi c}{L}\Big)
+\sum_{a\in A^{-}}f\Big(\frac{\rmi(c+\rmi a)}{L}\Big)
-\sum_{a\in A_{0}^{-}}f\Big(\frac{\rmi(c-\rmi a)}{L}\Big)\\
&&\hspace{62mm} +\sum_{a\in A^{+}}f\Big(\rho+\frac{\rmi(c-\rmi a)}{L}\Big)
-\sum_{a\in A_{0}^{+}}f\Big(\rho+\frac{\rmi(c+\rmi a)}{L}\Big)\;.\nonumber
\end{eqnarray}
The function $f$ verifies (\ref{f[fk]}) with first coefficients equal to $\alpha=\overline{\alpha}=1/2$, $f_{-1}=\overline{f}_{-1}=0$ and
\begin{equation}
f_{0}=\log\frac{(1+\rmi)\sqrt{2\pi}}{\sqrt{\rho(1-\rho)}}
\quad\text{and}\quad
\overline{f}_{0}=\log\frac{(1-\rmi)\sqrt{2\pi}}{\sqrt{\rho(1-\rho)}}\;.
\end{equation}
From (\ref{EM sqrt sqrt}), the expansion up to order $L^{0}$ of the sum over $j$ in the right hand side of (\ref{Xi2[A]}) is
\begin{eqnarray}
\label{tmp Xi2}
&& \sum_{j=1}^{N}f\Big(\frac{j-\frac{1}{2}+\rmi c}{L}\Big)
\simeq L\Big(\int_{0}^{\rho}\rmd u\,f(u)\Big)
+\log(2\pi)-\frac{\log\Gamma(\frac{1}{2}+\rmi c)}{2}-\frac{\log\Gamma(\frac{1}{2}-\rmi c)}{2}\\
&&\hspace{75mm} +f_{0}\,\zeta(0,\tfrac{1}{2}+\rmi c)
+\overline{f}_{0}\,\zeta(0,\tfrac{1}{2}-\rmi c)\;.\nonumber
\end{eqnarray}
Again, the integral can be computing as explained in appendix \ref{appendix integrals}. One finds $\int_{0}^{\rho}\rmd u\,f(u)=0$.

At leading order in $L$, using (\ref{Phi expansion}) and the constraint (\ref{m+-}) to treat the sums over $a$, and $\zeta(0,z)=\tfrac{1}{2}-z$ (\ref{Bernoulli[zeta]}), one can express the expansion of $\log \Xi_{2}$ to order $L^{0}$. After taking the exponential and using Euler's reflection formula $\Gamma(z)\Gamma(1-z)=\pi/\sin(\pi z)$ to eliminate the $\Gamma$ functions, we obtain
\begin{equation}
\label{asymptotics Xi2}
\prod_{j=1}^{N}\Big(1+\frac{1-\rho}{\rho}\,y_{j}\Big)\simeq\sqrt{1+\rme^{2\pi c}}\,\frac{\big(\prod_{a\in A^{+}}\sqrt{c-\rmi a}\big)\big(\prod_{a\in A^{-}}\sqrt{c+\rmi a}\big)}{\big(\prod_{a\in A_{0}^{+}}\sqrt{c+\rmi a}\big)\big(\prod_{a\in A_{0}^{-}}\sqrt{c-\rmi a}\big)}\;.
\end{equation}
The factor $\sqrt{1+\rme^{2\pi c}}$ has infinitely many branch cuts $\rmi(n+\half)+\mathbb{R}^{+}$, $n\in\mathbb{Z}$. Since $\log\Gamma$ is interpreted as the log $\Gamma$ function and not the logarithm of the $\Gamma$ function in (\ref{tmp Xi2}), $\sqrt{1+\rme^{2\pi c}}$ has to be understood for $\Re c>0$ as the analytic continuation in $c$ from the real axis $(-1)^{\big\lfloor\frac{\Im(4\pi c+2\rmi\pi)}{4\pi}\big\rfloor}\sqrt{1+\rme^{2\pi c}}$ with $\lfloor x\rfloor$ the largest integer smaller or equal to $x$. This corresponds to choosing instead the branch cuts $(-\rmi\infty,-\rmi/2]\cup[\rmi/2,\infty)$ for $\sqrt{1+\rme^{2\pi c}}$.
\end{subsection}

\begin{subsection}{Asymptotics of \texorpdfstring{$\Xi_{3}$}{Xi3}}
We consider the quantity $\Xi_{3}$, defined in (\ref{Xi3}). For the stationary state, one has $\Xi_{3}=1$. We focus on the other first eigenstates, and take the parameter $c$ with $\Re c>0$: this constraint is verified for the solution of (\ref{phi(c)}) as long as the real part of $s$ is not too negative. In particular, it seems to to be valid for all $s$ with $\Re s\geq0$. This is not the case for the stationary state, for which the solution of (\ref{phi(c)}) at leading order in $L$ is $c\to-\infty$ when $s\to0$.

Replacing $y_{j}$ and $y_{k}$ by $y_{j}^{0}$ and $y_{k}^{0}$ in the definition (\ref{Xi3}) of $\Xi_{3}$, with corrections coming from the sets $A_{0}^{\pm}$, $A^{\pm}$, one has
\begin{eqnarray}
\label{V/V0[Phi]}
&& \prod_{j=1}^{N}\prod_{k=j+1}^{N}\frac{y_{j}-y_{k}}{y_{j}^{0}-y_{k}^{0}}=\text{(finite products)}\\
&&\hspace{32mm}
  \times\frac
  {\prod\limits_{a\in A^{+}}\prod\limits_{j=1}^{N}\Big(\Phi\big(-\frac{\rho}{2}+\frac{j-\frac{1}{2}+\rmi c}{L}\big)-\Phi\big(\frac{\rho}{2}+\frac{a+\rmi c}{L}\big)\Big)}
  {\prod\limits_{a\in A_{0}^{+}}\prod\limits_{\substack{j=1\\n-j\neq a-\half}}^{N}\Big(\Phi\big(-\frac{\rho}{2}+\frac{j-\frac{1}{2}+\rmi c}{L}\big)-\Phi\big(\frac{\rho}{2}-\frac{a-\rmi c}{L}\big)\Big)}\nonumber\\
&&\hspace{32mm}
  \times
  \frac
  {\prod\limits_{a\in A^{-}}\prod\limits_{j=1}^{N}\Big(\Phi\big(-\frac{\rho}{2}+\frac{j-\frac{1}{2}+\rmi c}{L}\big)-\Phi\big(-\frac{\rho}{2}-\frac{a-\rmi c}{L}\big)\Big)}
  {\prod\limits_{a\in A_{0}^{-}}\prod\limits_{\substack{j=1\\j\neq a+\half}}^{N}\Big(\Phi\big(-\frac{\rho}{2}+\frac{j-\frac{1}{2}+\rmi c}{L}\big)-\Phi\big(-\frac{\rho}{2}+\frac{a+\rmi c}{L}\big)\Big)}\;.\nonumber
\end{eqnarray}
The $(\text{finite products})$ factor contains all the factors with only contributions from the sets $A_{0}^{\pm}$, $A^{\pm}$ and no product over $j$ between $1$ and $N$. Their asymptotics can be computed from (\ref{Phi expansion}) and (\ref{m+-}). At leading order in $L$, one finds
\begin{eqnarray}
&& (\text{finite products})\simeq\Big(\frac{(1-\rho)^{3/2}}{2\rmi\sqrt{\pi}\sqrt{\rho}}\,\sqrt{L}\Big)^{m_{r}}
  \Big(\prod_{a\in A_{0}^{+}}(-1)^{a-\frac{1}{2}}\Big)\Big(\prod_{a\in A_{0}^{-}}(-1)^{a+\frac{1}{2}}\Big)\\
&&\hspace{25mm}
  \times\prod_{\sigma\in\{-,+\}}\!\!\!
  \frac
  {\prod\limits_{\substack{a,a'\in A^{\sigma}\\\sigma a<\sigma a'}}\!\!\Big(\sqrt{c-\sigma\rmi a}-\sqrt{c-\sigma\rmi a'}\Big)\prod\limits_{\substack{a,a'\in A_{0}^{\sigma}\\\sigma a>\sigma a'}}\!\!\Big(\sqrt{c+\sigma\rmi a}-\sqrt{c+\sigma\rmi a'}\Big)}
  {\prod\limits_{a\in A^{\sigma}}\prod\limits_{a'\in A_{0}^{\sigma}}\!\!\Big(\sqrt{c-\sigma\rmi a}-\sqrt{c+\sigma\rmi a'}\Big)}
  \nonumber\\
&&\hspace{25mm}
  \times
  \frac
  {\prod\limits_{a\in A^{+}}\prod\limits_{a'\in A^{-}}\Big(\sqrt{c-\rmi a}+\sqrt{c+\rmi a'}\Big)\prod\limits_{a\in A_{0}^{+}}\prod\limits_{a'\in A_{0}^{-}}\Big(\sqrt{c+\rmi a}+\sqrt{c-\rmi a'}\Big)}
  {\prod\limits_{a\in A^{+}}\prod\limits_{a'\in A_{0}^{-}}\Big(\sqrt{c-\rmi a}+\sqrt{c-\rmi a'}\Big)\prod\limits_{a\in A_{0}^{+}}\prod\limits_{a'\in A^{-}}\Big(\sqrt{c+\rmi a}+\sqrt{c+\rmi a'}\Big)}
  \;.\nonumber
\end{eqnarray}

After taking the logarithm, the factors that contain a product over $j$ in (\ref{V/V0[Phi]}) take the form of a sum for $j$ between $1$ and $N$ of $\log f((j-\frac{1}{2}+\rmi c)/L)$ where
\begin{equation}
f(x)=\Phi\big(-\frac{\rho}{2}+x\big)-\Phi\big(\frac{\sigma\rho}{2}+\frac{\rmi c+\sigma'a}{L}\big)\;,
\end{equation}
with parameters, $\sigma,\sigma'\in\{+1,-1\}$, $a\in\mathbb{N}+\tfrac{1}{2}$.

The function $f$ has a singularity when either $x$ or $\rho-x$ is of order $1/L$: using (\ref{Phi expansion}) and the assumption $\Re c>0$, one has for large $L$
\begin{eqnarray}
f(x/L)&&\simeq\frac{\sqrt{2}(1+\rmi)\sqrt{\pi}\sqrt{\rho}}{(1-\rho)^{3/2}}\,\Big(\sqrt{\frac{x}{L}}+\sigma\sqrt{\frac{\rmi c+\sigma'a}{L}}\Big)\\
f(\rho-x/L)&&\simeq\frac{\sqrt{2}(1-\rmi)\sqrt{\pi}\sqrt{\rho}}{(1-\rho)^{3/2}}\,\Big(\sqrt{\frac{x}{L}}-\sigma\sqrt{-\frac{\rmi c+\sigma'a}{L}}\Big)\;.\nonumber
\end{eqnarray}
This is precisely the type of singularities that can be treated by (\ref{EM sqrt(Stirling)}). Since both ends of the summation range exhibit this singularity, one can directly use (\ref{EM sqrt(Stirling) sqrt(Stirling)}), with coefficients $\sigma_{0}=\sigma$, $\sigma_{1}=-\sigma$, $q_{0}=\rmi c+\sigma'a$, $q_{1}=-\rmi c-\sigma'a$. Introducing nonnegative integers $m$ and $m'$ to avoid terms in the sum for which $f((j-\frac{1}{2}+\rmi c)/L)=0$, we find the large $L$ asymptotics
\begin{eqnarray}
&& \sum_{j=m}^{N-m'}\log f\Big(\frac{j-\half+\rmi c}{L}\Big)
\simeq\rho\log\!\Big(\frac{\rho}{1-\rho}\Big)L+\sigma\frac{2\rmi\sqrt{\pi}\sqrt{\rho}\sqrt{c-\sigma'\rmi a}}{\sqrt{1-\rho}}\,\sqrt{L}+\frac{m+m'-1}{2}\,\log L\\
&&\hspace{35mm} +\log\sqrt{2\pi}+\frac{2\pi(1-2\rho)c}{3(1-\rho)}-\frac{\pi(1-5\rho)\sigma'\rmi a}{6(1-\rho)}
+\frac{\rmi\pi(1-m+m')}{4}\nonumber\\
&&\hspace{35mm} -(m+m'-1)\log\frac{2\sqrt{\pi}\sqrt{\rho}}{(1-\rho)^{3/2}}-\frac{\log\Gamma(m-\half-\sigma'a)}{2}-\frac{\log\Gamma(m'+\half+\sigma'a)}{2}\nonumber\\
&&\hspace{35mm} +\sigma\int_{\sigma'\rmi a}^{c}\rmd u\,\frac{\rme^{\rmi\pi/4}\zeta(\half,m-\half+\rmi u)-\rme^{-\rmi\pi/4}\zeta(\half,m'+\half-\rmi u)}{2\sqrt{u-\sigma'\rmi a}}\;.\nonumber
\end{eqnarray}
There, the integral giving the leading order in $L$ was computed as explained at the beginning of appendix \ref{appendix integrals} by making the change of variable $z=\Phi(u-\frac{\rho}{2})$. One has
\begin{eqnarray}
&& \int_{0}^{\rho}\rmd u\,\log f(u)=\rho\log\Big(-\Phi\Big(\frac{\sigma\rho}{2}+\frac{\rmi c+\sigma'a}{L}\Big)\Big)\\
&&\hspace{22mm} \simeq\rho\log\frac{\rho}{1-\rho}+\sigma\frac{2\rmi\sqrt{\pi}\sqrt{\rho}\sqrt{c-\sigma'\rmi a}}{\sqrt{1-\rho}\sqrt{L}}+\frac{2\pi(1-2\rho)(c-\sigma'\rmi a)}{3(1-\rho)L}\;.\nonumber
\end{eqnarray}

We write $\sum'$ for the full sum between $1$ and $N$ minus any divergent term as in (\ref{V/V0[Phi]}). Its expansion up to order $0$ in $L$ is
\begin{eqnarray}
\label{tmp Xi3}
&& \sum_{j=1}^{N}\!'\,\log f\Big(\frac{j-\half+\rmi c}{L}\Big)
\simeq\rho\log\!\Big(\frac{\rho}{1-\rho}\Big)L+\sigma\frac{2\rmi\sqrt{\pi}\sqrt{\rho}\sqrt{c-\sigma'\rmi a}}{\sqrt{1-\rho}}\,\sqrt{L}+\frac{1-\sigma\sigma'}{4}\,\log L\\
&&\hspace{30mm} +\log\sqrt{2\pi}+\frac{2\pi(1-2\rho)c}{3(1-\rho)}-\frac{\pi(1-5\rho)\sigma'\rmi a}{6(1-\rho)}+\frac{\rmi\pi(\sigma-\sigma')}{8}\nonumber\\
&&\hspace{30mm} -\frac{1-\sigma\sigma'}{2}\,\log\frac{2\sqrt{\pi}\sqrt{\rho}}{(1-\rho)^{3/2}}-\frac{\log\Gamma(m-\half-\sigma'a)}{2}-\frac{\log\Gamma(m'+\half+\sigma'a)}{2}\nonumber\\
&&\hspace{30mm} +\sigma\int_{\sigma'\rmi a}^{c}\rmd u\,\frac{\rme^{\rmi\pi/4}\zeta(\half,m-\half+\rmi u)-\rme^{-\rmi\pi/4}\zeta(\half,m'+\half-\rmi u)}{2\sqrt{u-\sigma'\rmi a}}\nonumber\\
&&\hspace{30mm} +\sum_{j=1}^{m-1}\!\!'\,\log\big(\sqrt{j-\half+\rmi c}+\sigma\sqrt{\rmi c+\sigma'a}\big)+\sum_{j=1}^{m'}\!'\,\log\big(\sqrt{j-\half-\rmi c}-\sigma\sqrt{-\rmi c-\sigma'a}\big)\;.\nonumber
\end{eqnarray}
The integers $m$ and $m'$ must be taken large enough so that the arguments of the $\Gamma$ functions do not belong to $-\mathbb{N}$. They are also needed to ensure the convergence of the integral at $u=\sigma'\rmi a$, since $\zeta(\half,-n+\varepsilon)\sim\varepsilon^{-1/2}$ for $n\in\mathbb{N}$. The branch cuts of the integrand as a function of $u$ are chosen equal to $(\rmi\infty,\rmi(m-\half)]$, $(-\rmi\infty,-\rmi(m'+\half)]$ and $(-\infty,\sigma'\rmi a]$.

Since the left hand side of (\ref{tmp Xi3}) is independent of $m$, $m'$, one would like to eliminate them also in the right hand side. This can be done using the relation
\begin{eqnarray}
\label{tmp identity Xi3 shift m}
&&\int_{-\Lambda}^{q}\rmd u\,\frac{\rme^{\rmi\pi/4}\zeta(\half,m-\half+\rmi u)-\rme^{-\rmi\pi/4}\zeta(\half,m'+\half-\rmi u)}{2\sqrt{u-\sigma'\rmi a}}\\
&& =\int_{-\Lambda}^{q}\rmd u\,\frac{\rme^{\rmi\pi/4}\zeta(\half,\half+\rmi u)-\rme^{-\rmi\pi/4}\zeta(\half,\half-\rmi u)}{2\sqrt{u-\sigma'\rmi a}}
-\frac{\rmi\pi(m+m'-1)}{4}\nonumber\\
&& -\sum_{j=1}^{m-1}\Big(\log\big(\sqrt{j-\half+\rmi q}+\sqrt{\sigma'a+\rmi q}\big)-\log\big(\sqrt{\Lambda+\rmi(j-\half)}+\sigma'\sqrt{\Lambda+\sigma'\rmi a}\big)\Big)\nonumber\\
&& +\sum_{j=1}^{m'}\Big(\log\big(\sqrt{j-\half-\rmi q}+\sqrt{-\sigma'a-\rmi q}\big)-\log\big(\sqrt{\Lambda-\rmi(j-\half)}-\sigma'\sqrt{\Lambda+\sigma'\rmi a}\big)\Big)\;,\nonumber
\end{eqnarray}
which follows from (\ref{sum_log_sqrt[Gamma,zeta]}) and is valid when $\Re q>0$ and $\Re\Lambda>0$ with $|\Im\Lambda|<a$ if the path of integration is required to avoid all branch cuts.

We decompose the integral from $\sigma'\rmi a$ to $c$ in (\ref{tmp Xi3}) as an integral from $-\Lambda$ to $c$ minus the limit $\varepsilon\to0$, $\varepsilon>0$ of an integral from $-\Lambda$ to $\sigma'\rmi a+\varepsilon$. After using (\ref{tmp identity Xi3 shift m}) for $q=c$ and $q=\sigma'\rmi a+\varepsilon$ the limit $\Lambda\to\infty$ of the integrals become convergent. The limit $\varepsilon\to0$ of the integral between $-\infty$ and $\sigma'\rmi a+\varepsilon$ (with a contour of integration that avoids the branch cut of $\zeta$) is given by the identity
\begin{equation}
\int_{-\infty}^{\sigma'\rmi a+\varepsilon}\rmd u\,\frac{\rme^{\rmi\pi/4}\zeta(\half,\half+\rmi u)-\rme^{-\rmi\pi/4}\zeta(\half,\half-\rmi u)}{2\sqrt{u-\sigma'\rmi a}}
\underset{\varepsilon\to0}{\simeq}\sigma'\log\sqrt{\varepsilon}+\sigma'\log\sqrt{8\pi}+\rmi\pi a\;,
\end{equation}
that was obtained numerically. The divergent contribution $\log\sqrt{\varepsilon}$ cancels with terms in the sums of (\ref{tmp identity Xi3 shift m}) at $j=a+\half$. After several simplifications, (\ref{tmp Xi3}) becomes
\begin{eqnarray}
&& \sum_{j=1}^{N}\!'\,\log f\Big(\frac{j-\half+\rmi c}{L}\Big)
\simeq\rho\log\!\Big(\frac{\rho}{1-\rho}\Big)L+\sigma\frac{2\rmi\sqrt{\pi}\sqrt{\rho}\sqrt{c-\sigma'\rmi a}}{\sqrt{1-\rho}}\,\sqrt{L}+\frac{1-\sigma\sigma'}{4}\,\log L\\
&&\hspace{34mm} -\frac{1-\sigma\sigma'}{2}\,\log\Big(\frac{\sqrt{\rho}}{2\sqrt{\pi}(1-\rho)^{3/2}\sqrt{c-\sigma'\rmi a}}\Big)+\frac{2\pi(1-2\rho)(c-\sigma'\rmi a)}{3(1-\rho)}\nonumber\\
&&\hspace{34mm} -(\sigma-\sigma')\rmi\pi(a-\tfrac{1}{4})+\sigma\int_{-\infty}^{c}\rmd u\,\frac{\rme^{\rmi\pi/4}\zeta(\half,\half+\rmi u)-\rme^{-\rmi\pi/4}\zeta(\half,\half-\rmi u)}{2\sqrt{u-\sigma'\rmi a}}\;.\nonumber
\end{eqnarray}
Putting everything together, we obtain for the product of the four factors of $\Xi_{3}$ containing products over $j$ in (\ref{V/V0[Phi]})
\begin{eqnarray}
&& \rme^{-\frac{2\rmi\sqrt{\pi}\sqrt{\rho}}{\sqrt{1-\rho}}\Big(\sum_{a\in A_{0}^{+}}\sqrt{c+\rmi a}+\sum_{a\in A^{-}}\sqrt{c+\rmi a}-\sum_{a\in A_{0}^{-}}\sqrt{c-\rmi a}-\sum_{a\in A^{+}}\sqrt{c-\rmi a}\Big)\,\sqrt{L}}\\
&& \times\Big(\frac{\sqrt{\rho}}{2\sqrt{\pi}(1-\rho)^{3/2}\sqrt{L}}\Big)^{m_{r}}
\rmi^{m_{r}^{+}-m_{r}^{-}}
\rme^{-\frac{2\rmi\pi(1-2\rho)p_{r}}{3(1-\rho)}}
\Big(\prod_{a\in A_{0}^{+}}\frac{1}{\sqrt{c+\rmi a}}\Big)
\Big(\prod_{a\in A_{0}^{-}}\frac{1}{\sqrt{c-\rmi a}}\Big)\nonumber\\
&& \times\exp\Bigg[-\frac{1}{2}\int_{-\infty}^{c}\rmd u\,\Big(\rme^{\rmi\pi/4}\zeta(\half,\half+\rmi u)-\rme^{-\rmi\pi/4}\zeta(\half,\half-\rmi u)\Big)\nonumber\\
&&\hspace{25mm} \times\bigg(\sum_{a\in A_{0}^{+}}\frac{1}{\sqrt{u+\rmi a}}+\sum_{a\in A^{-}}\frac{1}{\sqrt{u+\rmi a}}-\sum_{a\in A_{0}^{-}}\frac{1}{\sqrt{u-\rmi a}}-\sum_{a\in A^{+}}\frac{1}{\sqrt{u-\rmi a}}\bigg)\Bigg]\;.\nonumber
\end{eqnarray}
\end{subsection}

\begin{subsection}{Asymptotics of \texorpdfstring{$\Xi_{4}$}{Xi4}}
We write
\begin{equation}
\label{Xi4[sum]}
\prod_{j=1}^{N}\prod_{j'=j+1}^{N}\big(y_{j}^{0}-y_{j'}^{0}\big)=\rmi^{\frac{N(N-1)}{2}}\exp\bigg(\sum_{j=1}^{N}\sum_{j'=j+1}^{N}\log(-\rmi(y_{j}^{0}-y_{j'}^{0}))\bigg)\;,
\end{equation}
where $\log$ is the usual determination of the logarithm with branch cut $\mathbb{R}^{-}$. The (clockwise) contour on which the $y_{j}^{0}$'s condense in the complex plane is represented in figure \ref{fig contour Phi}. The factor $-\rmi$ in the logarithm ensures that the branch cut of the logarithm is not crossed.

Since $y_{j}^{0}\to-\rho/(1-\rho)$ when $j\ll L$ and when $N-j\ll L$, the quantity $y_{j}^{0}-y_{j'}^{0}$ goes to $0$ at the three corners of the triangle $\{(j,j'),1\leq j<j'\leq N\}$. To avoid extra singularities and allow us to use (\ref{EM triangle sqrt sqrt sqrt}), we consider first the regular part
\begin{equation}
S_{\text{reg}}=\sum_{j=1}^{N}\sum_{j'=j+1}^{N}f\Big(\frac{j-\half+\rmi c}{L},\frac{j'-\half+\rmi c}{L}\Big)\;
\end{equation}
with $f$ defined by
\begin{equation}
f(u,v)=\log\bigg(\!-\rmi\Big(\Phi\big(-\frac{\rho}{2}+u\big)-\Phi\big(-\frac{\rho}{2}+v\big)\Big)\,\frac{(\sqrt{u}+\sqrt{v})(\sqrt{\rho-u}+\sqrt{\rho-v})}{(v-u)(\sqrt{-\rmi u}+\sqrt{\rmi(\rho-v)})}\bigg).
\end{equation}
Using the same notations as in (\ref{EM triangle sqrt sqrt sqrt}), one has
\begin{eqnarray}
&& f_{0,0}=\log\frac{4\rmi\sqrt{\pi}\sqrt{\rho}}{(1-\rho)^{3/2}}\;,\qquad
g_{0,0}=\log\Big(-\frac{4\rmi\sqrt{\pi}\sqrt{\rho}}{(1-\rho)^{3/2}}\Big)\;,\qquad
h_{0,0}=\log\Big(\frac{2\sqrt{\pi}\sqrt{\rho}}{(1-\rho)^{3/2}}\Big)\;,\\
&& f_{1}(v)=\frac{1}{\sqrt{v}}+\frac{\rmi}{\sqrt{\rho-v}}-\frac{2\sqrt{\rmi}\sqrt{\pi}\sqrt{\rho}}{\sqrt{1-\rho}(\rho+(1-\rho)\Phi(v-\rho/2))}\;,\nonumber\\
&& g_{1}(v)=\frac{1}{\sqrt{v}}-\frac{\rmi}{\sqrt{\rho-v}}-\frac{2\sqrt{-\rmi}\sqrt{\pi}\sqrt{\rho}}{\sqrt{1-\rho}(\rho+(1-\rho)\Phi(\rho/2-v))}\;,\nonumber\\
&& f_{2}(v)=-\frac{1}{2(\rho-v)}+\frac{\sqrt{\rho-v}}{2\sqrt{\rho}\,v}-\frac{2\rmi\pi\rho}{(1-\rho)(\rho+(1-\rho)\Phi(v-\rho/2))^{2}}+\frac{4\rmi\pi(1+\rho)}{3(1-\rho)(\rho+(1-\rho)\Phi(v-\rho/2))}\;,\nonumber\\
&& g_{2}(v)=-\frac{1}{2(\rho-v)}+\frac{\sqrt{\rho-v}}{2\sqrt{\rho}\,v}+\frac{2\rmi\pi\rho}{(1-\rho)(\rho+(1-\rho)\Phi(\rho/2-v))^{2}}-\frac{4\rmi\pi(1+\rho)}{3(1-\rho)(\rho+(1-\rho)\Phi(\rho/2-v))}\;.\nonumber
\end{eqnarray}
When the two arguments of $f$ are equal, one finds the limits
\begin{eqnarray}
&& f(v,v)=\log\Big(\frac{8\pi\sqrt{v}\sqrt{\rho-v}\,\Phi(v-\rho/2)(1-\Phi(v-\rho/2))}{(\sqrt{-\rmi v}+\sqrt{\rmi(\rho-v)})(\rho+(1-\rho)\Phi(v-\rho/2))}\Big)\;,\\
&& f^{(1,0)}(v,v)=-\frac{1}{2\rho}+\frac{\rmi\pi}{1-\rho}+\frac{1}{4v}-\frac{1}{4(\rho-v)}+\frac{\rmi\sqrt{\rho-v}}{2\rho\sqrt{v}}-\frac{\rmi\pi\rho}{(1-\rho)(\rho+(1-\rho)\Phi(v-\rho/2))^{2}}\;,\nonumber\\
&& f^{(0,1)}(v,v)=\frac{1}{2\rho}+\frac{\rmi\pi}{1-\rho}+\frac{1}{4v}-\frac{1}{4(\rho-v)}+\frac{\rmi\sqrt{v}}{2\rho\sqrt{\rho-v}}-\frac{\rmi\pi\rho}{(1-\rho)(\rho+(1-\rho)\Phi(v-\rho/2))^{2}}\;.\nonumber
\end{eqnarray}
One can use (\ref{EM triangle sqrt sqrt sqrt}) for the asymptotic expansion. After some simplifications, which involve in particular the calculation of several simple and double integrals explained in appendix \ref{appendix integrals}, and using the explicit values (\ref{Bernoulli[zeta]}) and (\ref{Hurwitz zeta double 00}) for $\zeta(0,z)$, $\zeta(-1,z)$ and $\tilde{\zeta}_{0}(0,0,z,z')$, one finds
\begin{eqnarray}
&& S_{\text{reg}}\simeq
L^{2}\Big(\frac{9\rho^{2}}{8}+\frac{\rho^{2}\log\rho}{4}+\frac{\rho(1-\rho)\log(1-\rho)}{2}\Big)\\
&&\hspace{7mm} +L\Big(-\frac{\rho\log\rho}{4}-\frac{(1-\rho)\log(1-\rho)}{2}+\frac{\rho}{2}-\frac{\rho\log(8\pi)}{2}+\frac{\pi\rho c}{2}\Big)\nonumber\\
&&\hspace{7mm} +\sqrt{L}\Big(2\sqrt{\rho}-\frac{\sqrt{2\pi}\sqrt{\rho}}{\sqrt{1-\rho}}\Big)\Big((1+\rmi)\zeta(-\half,\half+\rmi c)+(1-\rmi)\zeta(-\half,\half-\rmi c)\Big)
+\Big(-\frac{1}{8}+\frac{\log2}{12}-\frac{\pi c}{2}-c^{2}\Big)\;.\nonumber
\end{eqnarray}

The singular part $S_{\text{sng}}$ of the sum in (\ref{Xi4[sum]}) is equal to
\begin{eqnarray}
\label{tmp Xi4 singular part}
&& \sum_{j=1}^{N}\sum_{j'=j+1}^{N}\log\frac{\Big(\frac{j'-j}{L}\Big)\Big(\sqrt{-\rmi\,\frac{j-\half+\rmi c}{L}}+\sqrt{\rmi\,\frac{j'-\half+\rmi c}{L}}\Big)}{\Big(\sqrt{\frac{j-\half+\rmi c}{L}}+\sqrt{\frac{j'-\half+\rmi c}{L}}\Big)\Big(\sqrt{\rho-\frac{j-\half+\rmi c}{L}}+\sqrt{\rho-\frac{j'-\half+\rmi c}{L}}\Big)}\\
&& =N\log2-\frac{N(N-1)}{4}\log L+\log(G(N+1))+\frac{1}{4}\log\frac{\Gamma(N+\half+\rmi c)\Gamma(N+\half-\rmi c)}{\Gamma(\half+\rmi c)\Gamma(\half-\rmi c)}\nonumber\\
&& +\sum_{j=1}^{N}\sum_{j'=1}^{N}\log\big(\sqrt{-\rmi(j-\half+\rmi c)}+\sqrt{\rmi(j'-\half-\rmi c)}\big)\nonumber\\
&& -\sum_{j=1}^{N}\sum_{j'=1}^{N}1_{\{j+j'>N\}}\log\big(\sqrt{-\rmi(j-\half+\rmi c)}+\sqrt{\rmi(j'-\half-\rmi c)}\big)\nonumber\\
&& -\frac{1}{2}\sum_{j=1}^{N}\sum_{j'=1}^{N}\log(\sqrt{j-\half+\rmi c}+\sqrt{j'-\half+\rmi c})
-\frac{1}{2}\sum_{j=1}^{N}\sum_{j'=1}^{N}\log(\sqrt{j-\half-\rmi c}+\sqrt{j'-\half-\rmi c})\;,\nonumber
\end{eqnarray}
where $G$ is Barnes function $G(N+1)=\prod_{j=1}^{N-1}j!$, whose large $N$ expansion is given by (\ref{Barnes G asymptotics}). The expansions of the remaining sums are obtained from (\ref{EM rectangle log(sqrt+sqrt) i}), (\ref{EM rectangle log(sqrt+sqrt)}) and
\begin{eqnarray}
&& \sum_{j=1}^{N}\sum_{j'=1}^{N}1_{\{j+j'>N\}}\log\Big(\sqrt{-\rmi(j-\half+\rmi c)}+\sqrt{\rmi(j'-\half-\rmi c)}\Big)
\simeq\frac{N^{2}\log N}{4}+\Big(\log2-\frac{5}{8}\Big)N^{2}\\
&&\hspace{85mm} +\frac{N\log N}{4}-\frac{(4-\pi)c}{2}\,N-\Big(\frac{1+\log2}{24}-\frac{\pi c}{4}\Big)\;,\nonumber
\end{eqnarray}
which can be derived by cutting the triangle into two triangles and a rectangle and using (\ref{EM rectangle sqrt}) and (\ref{EM triangle sqrt}).

The expansions (\ref{EM rectangle log(sqrt+sqrt) i}) and (\ref{EM rectangle log(sqrt+sqrt)}) contribute the constants $\kappa_{1}(-1)$ and $\kappa_{0}$ as $\kappa_{1}(-1)-\kappa(0)\approx$0.1819507467 3927841. These constants can be evaluated numerically with very large precision using BST algorithm as described in section \ref{section numerics}. From numerical computations, we conjecture the identity
\begin{equation}
\kappa_{1}(-1)-\kappa(0)=\frac{1}{12}-\zeta'(-1)-\frac{\log2}{8}-\int_{-\infty}^{0}\rmd u\,\Big(\rme^{\rmi\pi/4}\zeta(\half,\half+\rmi u)-\rme^{-\rmi\pi/4}\zeta(\half,\half-\rmi u)\Big)^{2}\;,
\end{equation}
that was checked within $100$ significant digits. Our derivation of the asymptotics (\ref{asymptotics norm}) of the norm $\mathcal{N}_{r}(\gamma)$ does not rely heavily on this numerical conjecture since the value of $\kappa_{1}(-1)-\kappa(0)$ can in fact also be inferred from the stationary value $\mathcal{N}_{0}(0)=1$.

Gathering the various contributions to the singular term, one finds
\begin{eqnarray}
&& S_{\text{sng}}\simeq\Big(\frac{\log\rho}{4}-\frac{9}{8}\Big)\rho^{2}L^{2}+\frac{\rho L\log L}{2}+\Big(\frac{\log\rho}{4}-\frac{1}{2}+\frac{\log(8\pi)}{2}+\frac{\pi c}{2}\Big)\rho L\nonumber\\
&&\hspace{10mm} -2\sqrt{\rho}\sqrt{L}\Big((1+\rmi)\zeta(-\half,\half+\rmi c)+(1-\rmi)\zeta(-\half,\half-\rmi c)\Big)\nonumber\\
&&\hspace{10mm} +\frac{1}{8}-\frac{\log2}{12}+\frac{\log(2\pi)}{4}-\frac{\pi c}{4}+c^{2}-\frac{\log\Gamma(\half+\rmi c)}{4}-\frac{\log\Gamma(\half-\rmi c)}{4}\nonumber\\
&&\hspace{10mm} -\frac{1}{4}\int_{-\infty}^{c}\rmd u\,\Big(\rme^{\rmi\pi/4}\zeta(\half,\half+\rmi u)-\rme^{-\rmi\pi/4}\zeta(\half,\half-\rmi u)\Big)^{2}\;.
\end{eqnarray}

Putting the regular and the singular terms together, taking the exponential, and using Euler's reflection formula $\Gamma(z)\Gamma(1-z)=\pi/\sin(\pi z)$ to eliminate the $\Gamma$ functions, we finally obtain
\begin{eqnarray}
&& \prod_{j=1}^{N}\prod_{j'=j+1}^{N}(y_{j}^{0}-y_{j'}^{0})
\simeq\rmi^{\frac{N(N-1)}{2}}\rme^{\frac{\rho bL^{2}}{2}}\rme^{\frac{\rho L\log L}{2}}\rme^{-\frac{(1-\rho)\log(1-\rho)L}{2}}\\
&&\hspace{30mm} \times\exp\Big(-\frac{2\sqrt{\pi}\sqrt{\rho}}{\sqrt{1-\rho}}\big(\rme^{\rmi\pi/4}\zeta(-\half,\half+\rmi c)+\rme^{-\rmi\pi/4}\zeta(-\half,\half-\rmi c)\big)\sqrt{L}\Big)\nonumber\\
&&\hspace{30mm} \times\rme^{-\pi c}(1+\rme^{2\pi c})^{1/4}
\,\exp\Bigg[-\frac{1}{4}\int_{-\infty}^{c}\rmd u\,\Big(\rme^{\rmi\pi/4}\zeta(\half,\half+\rmi u)-\rme^{-\rmi\pi/4}\zeta(\half,\half-\rmi u)\Big)^{2}\Bigg]\;,\nonumber
\end{eqnarray}
where $(1+\rme^{2\pi c})^{1/4}$ has to be interpreted once again as the analytic continuation in $c$ from the real axis as explained after (\ref{asymptotics Xi2}).
\end{subsection}

\begin{subsection}{Asymptotics of \texorpdfstring{$\Xi_{1}^{-1}\Xi_{2}^{-1}\Xi_{3}^{2}\,\Xi_{4}^{2}$}{(Xi3^2*Xi4^2)/(Xi1*Xi2)}}
We observe that many simplifications occur when multiplying $\Xi_{3}$ and $\Xi_{4}$ if we replace the $\zeta$ function by the eigenstate-dependent function $\varphi_{r}$ defined in (\ref{phi[A,zeta]}). One has
\begin{eqnarray}
&& \prod_{j=1}^{N}\prod_{k=j+1}^{N}(y_{j}-y_{k})
\simeq\rmi^{\frac{N(N-1)}{2}}
\rme^{\frac{\rho b L^{2}}{2}}
\rme^{\frac{\rho L \log L}{2}}
\rme^{-\frac{(1-\rho)\log(1-\rho)L}{2}}
\rme^{-\frac{\sqrt{\rho}\varphi_{r}(2\pi c)\sqrt{L}}{\sqrt{1-\rho}}}
\rme^{-\frac{2\rmi\pi(1-2\rho)p_{r}}{3(1-\rho)}}\\
&&\hspace{5mm} \frac{(\pi/2)^{m_{r}^{2}}}{(2\pi)^{m_{r}}}
\,\rme^{\rmi\pi\big(\sum_{a\in A_{0}^{+}}a-\sum_{a\in A_{0}^{-}}a\big)}
\omega(A_{0}^{+})\omega(A_{0}^{-})\omega(A^{+})\omega(A^{-})\omega(A_{0}^{+},A_{0}^{-})\omega(A^{+},A^{-})\nonumber\\
&&\hspace{5mm} \frac{(1+\rme^{2\pi c})^{1/4}}{\rme^{\pi c}}
\frac{\big(\prod_{a\in A^{+}}(c-\rmi a)^{1/4}\big)\big(\prod_{a\in A^{-}}(c+\rmi a)^{1/4}\big)}{\big(\prod_{a\in A_{0}^{+}}(c+\rmi a)^{1/4}\big)\big(\prod_{a\in A_{0}^{-}}(c-\rmi a)^{1/4}\big)}
\,\exp\!\Big(\lim_{\Lambda\to\infty}-m_{r}^{2}\log\Lambda+\int_{-\Lambda}^{2\pi c}\rmd u\,\frac{(\varphi_{r}'(u))^{2}}{2}\Big)
\;,\nonumber
\end{eqnarray}
where the combinatorial factors $\omega$ are defined in (\ref{omega(A)}). The limit $\Lambda\to\infty$ is needed to define the integral, which is divergent when $u=-\infty$ (except for the stationary state $m_{r}^{+}=m_{r}^{-}=0$) since $\varphi_{r}'(u)\sim u^{-1/2}$ when $u\to-\infty$. Even more simplifications occur after dividing $\Xi_{3}^{2}\Xi_{4}^{2}$ by $\Xi_{1}\Xi_{2}$. One finally finds
\begin{eqnarray}
\label{asymptotics (Xi3*Xi4)/(Xi1*Xi2)}
&& \frac{\prod_{j=1}^{N}\prod_{k=j+1}^{N}(y_{j}-y_{k})^{2}}
{\Big(\frac{1}{N}\sum_{j=1}^{N}\frac{y_{j}}{\rho+(1-\rho)y_{j}}\Big)\Big(\prod_{j=1}^{N}\Big(1+\frac{1-\rho}{\rho}\,y_{j}\Big)\Big)}
\simeq
\rme^{\rho\,bL^{2}}\rme^{\rho\,L\log L}\rme^{-(1-\rho)\log(1-\rho)L}\,\exp\Big(\!-\frac{2\sqrt{\rho}\,\varphi_{r}(2\pi c)}{\sqrt{1-\rho}}\,\sqrt{L}\Big)\nonumber\\
&&\hspace{4mm} \times\frac{(-1)^{\frac{N(N-1)}{2}+m_{r}}}{(\pi^{2}/4)^{-m_{r}^{2}}(4\pi^{2})^{m_{r}}}\,\exp\Big(-\frac{4\rmi\pi(1-2\rho)p_{r}}{3(1-\rho)}\Big)\,\omega(A_{0}^{+})^{2}\omega(A_{0}^{-})^{2}\omega(A^{+})^{2}\omega(A^{-})^{2}\omega(A_{0}^{+},A_{0}^{-})^{2}\omega(A^{+},A^{-})^{2}\nonumber\\
&&\hspace{4mm} \times\frac{\sqrt{\rho(1-\rho)}\sqrt{L}}{\rme^{2\pi c}\,\varphi_{r}'(2\pi c)}\,\exp\!\Big(\lim_{\Lambda\to\infty}-2m_{r}^{2}\log\Lambda+\int_{-\Lambda}^{2\pi c}\rmd u\,(\varphi_{r}'(u))^{2}\Big)\;.
\end{eqnarray}
We observe that several factors depending on $c$ have cancelled: the only dependency on $c$ left are $\varphi_{r}(2\pi c)$, $\rme^{2\pi c}\varphi_{r}'(2\pi c)$, and the upper limit of the integral.
\end{subsection}
\end{section}

\begin{section}{Conclusions}
It was shown in \cite{P2014.1} that the first eigenvalues of TASEP are naturally expressed in terms of a function $\eta$ constructed from the elementary excitations characterizing the eigenstate. We extend that result here by showing that $\varphi=\eta'$ can be identified as a scalar field: indeed, we observe that the normalization of the corresponding Bethe eigenstates can be expressed in terms of the exponential of the free action of $\varphi$. This might hint at a field theoretic description of current fluctuations on the relaxation scale $T\sim L^{3/2}$.

The field $\varphi$ is equal to a sum of square roots corresponding to elementary excitations over a Fermi sea, plus Hurwitz zeta functions corresponding to a kind of renormalization of the contribution of the Fermi sea. In our calculations, the contributions leading to these two parts of the field need to be treated separately. It is only at the end of the calculation that everything combines perfectly at several places to give exactly the same field $\varphi$ everywhere. It would be very nice to find a simpler derivation that makes it clearer why the field $\varphi$ should appears in the end, and to explain all the other unexpected cancellations that happen between the asymptotic expansions of seemingly very different factors.
\end{section}

\appendix
\normalsize
\begin{section}{Hurwitz zeta function and double Hurwitz zeta function}
\label{appendix zeta}
In this appendix, we summarize some properties of Hurwitz zeta function and double Hurwitz zeta function.

\begin{subsection}{Hurwitz zeta function}

\begin{subsubsection}{Definitions}
Hurwitz zeta function is defined for $\Re s>1$ by
\begin{equation}
\label{Hurwitz zeta}
\zeta(s,z)=\sum_{j=0}^{\infty}(j+z)^{-s}\;.
\end{equation}
For $z\not\in\mathbb{R}^{-}$, it can be analytically continued to a meromorphic function of $s$ with a simple pole at $s=1$ with residue equal to $1$, independent of $z$. It is convenient for using the Euler-Maclaurin formula to define a modification $\tilde{\zeta}$ of $\zeta$ such that $\tilde{\zeta}(s,z)=\zeta(s,z)$ when $s\neq1$ and
\begin{equation}
\label{Hurwitz zeta tilde}
\tilde{\zeta}(1,z)=\lim_{s\to1}\zeta(s,z)-\frac{1}{s-1}=-\frac{\Gamma'(z)}{\Gamma(z)}\;.
\end{equation}
The modified function $\tilde{\zeta}$ is not continuous at $s=1$. It obeys however the property
\begin{equation}
\lim_{s\to1}\big(\zeta(s,z)-\zeta(s,z')\big)=\tilde{\zeta}(s,z)-\tilde{\zeta}(s,z')\;.
\end{equation}
\end{subsubsection}

\begin{subsubsection}{Derivative}
Hurwitz zeta function verifies
\begin{equation}
\label{zeta'}
\partial_{z}\zeta(s,z)=-s\zeta(s+1,z)\;.
\end{equation}
\end{subsubsection}

\begin{subsubsection}{Bernoulli polynomials}
Hurwitz zeta function is related to Bernoulli polynomials. For $r\in\mathbb{N}^{*}$, one has
\begin{equation}
\label{Bernoulli[zeta]}
B_{r}(z)=-r\zeta(1-r,z)\;,
\end{equation}
which is a polynomial in $z$ of degree $r$. The Bernoulli polynomials form an Appell sequence, \textit{i.e.} $B'_{r}(x)=rB_{r-1}(x)$, or equivalently
\begin{equation}
\label{Bernoulli sum}
B_{r}(x+y)=\sum_{m=0}^{r}\C{r}{m}B_{m}(x)y^{r-m}\;.
\end{equation}
They also verify the symmetry relation
\begin{equation}
\label{Bernoulli symmetry}
B_{r}(x+1)=(-1)^{r}B_{r}(-x)\;.
\end{equation}
\end{subsubsection}

\begin{subsubsection}{Asymptotic expansions}
When its second argument becomes large, Hurwitz zeta has the asymptotic expansion
\begin{equation}
\label{Hurwitz zeta asymptotics}
\zeta(s,M+x)
\simeq-\frac{1}{1-s}\sum_{\ell=0}^{\infty}\C{1-s}{\ell}\frac{B_{\ell}(x)}{M^{\ell+s-1}}\;,
\end{equation}
while for $\tilde{\zeta}$, Stirling's formula leads to
\begin{equation}
\label{Hurwitz zeta tilde asymptotics}
\tilde{\zeta}(1,M+x)\simeq-\log M+\sum_{\ell=1}^{\infty}\frac{(-1)^{\ell}}{\ell}\,\frac{B_{\ell}(x)}{M^{\ell}}\;.
\end{equation}
\end{subsubsection}

\end{subsection}

\begin{subsection}{Double Hurwitz zeta function}
A two-dimensional generalization, the double Hurwitz zeta function, can be defined as
\begin{equation}
\label{Hurwitz zeta double}
\zeta(s,s';z,z')=\sum_{j=0}^{\infty}\sum_{j'=j+1}^{\infty}(j+z)^{-s}(j'+z')^{-s'}\;.
\end{equation}
The sum converges for $z$ and $z'$ outside $\mathbb{R}^{-}$ when $\Re s'>1$ and $\Re(s+s')>2$. Unlike the usual Hurwitz zeta function, there does not seem to be a standard accepted notation here, partly due to the fact that several natural two dimensional generalizations can be considered.

\begin{subsubsection}{Analytic continuation}
The analytic continuation to arbitrary $s$, $s'$ can be made \cite{M2002.1} using the Mellin-Barnes integral formula, which can be stated for $-\Re s<p<0$, $\lambda\not\in\mathbb{R}^{-}$ as
\begin{equation}
\label{Mellin-Barnes}
(1+\lambda)^{-s}=\int_{p-\rmi\infty}^{p+\rmi\infty}\frac{\rmd w}{2\rmi\pi}\,\frac{\Gamma(s+w)\Gamma(-w)}{\Gamma(s)}\,\lambda^{w}\;,
\end{equation}
and which follows from closing the contour of integration on the right and calculating the residues on the positive real axis when $|\lambda|<1$. Rewriting (\ref{Hurwitz zeta double}) as
\begin{equation}
\zeta(s,s';z,z')
=\sum_{j=0}^{\infty}\sum_{j'=0}^{\infty}(j+z)^{-s}(j'+z'-z+1)^{-s'}\Big(1+\frac{j+z}{j'+z'-z+1}\Big)^{-s'}\;,
\end{equation}
applying (\ref{Mellin-Barnes}) to the factor $(1+\tfrac{j+z}{j'+z'-z+1})^{-s'}$, shifting $w$ by $-s'$, and using (\ref{sum_power[zeta]}) to compute the sums over $j$ and $j'$ (provided $1<\Re w<\Re(s+s'-1)$) leads to
\begin{equation}
\zeta(s,s';z,z')
=\int_{p-\rmi\infty}^{p+\rmi\infty}\frac{\rmd w}{2\rmi\pi}\,\frac{\Gamma(w)\Gamma(s'-w)\zeta(s+s'-w,z)\zeta(w,z'-z+1)}{\Gamma(s')}\;
\end{equation}
with $1<p<\Re s'$. The contour of integration can be shifted to $0<p<1$ by taking the residue coming from the simple pole with residue $1$ of $\zeta(w,z'-z+1)$ at $w=1$. Shifting again the contour of integration to the left we pick the residues at $w=-k$, $k\in\mathbb{N}$ coming from $\Gamma(w)$ (residue $(-1)^{k}/k!$ at $w=-k$). Shifting $k$ by $1$, one finally finds in terms of Bernoulli polynomials (\ref{Bernoulli[zeta]})
\begin{eqnarray}
\label{Hurwitz zeta double sum integral}
&& \zeta(s,s';z,z')=\frac{1}{s'-1}\sum_{\ell=0}^{m+1}\C{1-s'}{\ell}\zeta(s+s'+\ell-1,z)B_{\ell}(z'-z+1)\nonumber\\
&&\hspace{20mm} +\int_{-m-\half-\rmi\infty}^{-m-\half+\rmi\infty}\frac{\rmd w}{2\rmi\pi}\,\frac{\Gamma(w)\Gamma(s'-w)\zeta(s+s'-w,z)\zeta(w,z'-z+1)}{\Gamma(s')}\;.
\end{eqnarray}
The remaining integral is analytic in the domain $\{(s,s'),\Re s'>-m-\half,\Re(s+s')>-m+\half\}$. It implies that $\zeta(s,s';z,z')$ is a meromorphic function of $s$, $s'$ with (possible) poles at $s'=1$ and $s+s'=2-n$, $n\in\mathbb{N}$. When approaching the pole at $s+s'=2-n$, one has (when $s,s'\not\in1-\mathbb{N}$)
\begin{eqnarray}
&& \zeta(s+\alpha\varepsilon,s'+(1-\alpha)\varepsilon;z,z')
\underset{\varepsilon\to0}{\simeq}\frac{1}{s'-1}\C{1-s'}{n}B_{n}(z'-z+1)\\
&&\hspace{50mm} \times\Big(\frac{1}{\varepsilon}-\frac{\Gamma'(z)}{\Gamma(z)}-(1-\alpha)\frac{\Gamma'(1-s')}{\Gamma(1-s')}+(1-\alpha)\frac{\Gamma'(s)}{\Gamma(s)}\Big)\;.\nonumber
\end{eqnarray}
The arbitrary parameter $\alpha$ characterizes the direction in which $(s,s')$ approaches the line $s+s'=2-n$. We define a modified version $\tilde{\zeta}_{\alpha}$ of double Hurwitz zeta, equal to $\zeta$ when $2-s-s'\not\in\mathbb{N}$, and to
\begin{equation}
\tilde{\zeta}_{\alpha}(s,s';z,z')=\lim_{\varepsilon\to0}\Big(\zeta(s+\alpha\varepsilon,s'+(1-\alpha)\varepsilon;z,z')
-\frac{1}{s'-1}\C{1-s'}{n}\frac{B_{n}(z'-z+1)}{\varepsilon}\Big)\;
\end{equation}
when $s+s'=2-n$, $n\in\mathbb{N}$. More explicitly
\begin{eqnarray}
\label{Hurwitz zeta double tilde sum integral}
&& \tilde{\zeta}_{\alpha}(s,s';z,z')=\frac{1}{s'-1}\sum_{\substack{\ell=0\\(\ell\neq n)}}^{m+1}\C{1-s'}{\ell}\zeta(\ell-n+1,z)B_{\ell}(z'-z+1)\\
&&\hspace{20mm} -\frac{1}{s'-1}\C{1-s'}{n}B_{n}(z'-z+1)\bigg(\frac{\Gamma'(z)}{\Gamma(z)}+(1-\alpha)\Big(\frac{\Gamma'(1-s')}{\Gamma(1-s')}-\frac{\Gamma'(s)}{\Gamma(s)}\Big)\bigg)\nonumber\\
&&\hspace{20mm} +\int_{-m-\half-\rmi\infty}^{-m-\half+\rmi\infty}\frac{\rmd w}{2\rmi\pi}\,\frac{\Gamma(w)\Gamma(s'-w)\zeta(s+s'-w,z)\zeta(w,z'-z+1)}{\Gamma(s')}\;,\nonumber
\end{eqnarray}
with $m\geq n-1$, $m\geq-\Re s'-\half$ and when $s,s'\not\in1-\mathbb{N}$. For $\alpha=1$, this corresponds to replacing the simple Hurwitz zeta function in the summation in (\ref{Hurwitz zeta double sum integral}) by its modified value (\ref{Hurwitz zeta tilde}).
\end{subsubsection}

\begin{subsubsection}{Double Bernoulli polynomials}
The modified double zeta functions can be extended to $s\in-\mathbb{N}$, $s'\in-\mathbb{N}$. There, the integral vanishes because of the $\Gamma(s')$ in the denominator and $\tilde{\zeta}_{\alpha}(s,s';z,z')$ become polynomials in $z$, $z'$. Using (\ref{Bernoulli[zeta]}), we obtain ($s+s'=2-n$)
\begin{equation}
\label{Hurwitz zeta double integers}
\tilde{\zeta}_{\alpha}(s,s';z,z')=\frac{1}{s'-1}\sum_{\ell=0}^{n-1}\C{1-s'}{\ell}\frac{B_{\ell}(z'-z+1)B_{n-\ell}(z)}{\ell-n}
+B_{n}(z'-z+1)\frac{(-1)^{s}(1-\alpha)}{(1-s)(1-s')\C{2-s-s'}{1-s}}\;.
\end{equation}
In particular, at $s=s'=0$, one has
\begin{equation}
\label{Hurwitz zeta double 00}
\tilde{\zeta}_{\alpha}(0,0;z,z')=-\frac{1+\alpha}{12}+\frac{\alpha(z-z')}{2}-\frac{\alpha z^{2}}{2}+\frac{(1-\alpha)(z')^{2}}{2}+\alpha zz'\;.
\end{equation}
Unlike the one-dimensional case, there is no unique natural way to define double Bernoulli numbers and polynomials because of the arbitrary parameter $\alpha$.
\end{subsubsection}

\begin{subsubsection}{Asymptotic expansion}
The expressions (\ref{Hurwitz zeta double sum integral}) gives the large $M$ asymptotic expansion of $\zeta(s,s';z+M,z'+M)$ using the one (\ref{Hurwitz zeta asymptotics}) for simple Hurwitz zeta since $\zeta(s+s'-w,M+z)\sim M^{-m+\half-s-s'}$ in the integral can be made arbitrarily small by taking $m$ large enough. This gives the asymptotic expansion
\begin{equation}
\label{Hurwitz zeta double asymptotics}
\zeta(s,s';M+z,M+z')\simeq\sum_{\ell=0}^{\infty}\sum_{m=0}^{\infty}\frac{\C{1-s'}{\ell}}{1-s'}\,\frac{\C{2-s-s'-\ell}{m}}{2-s-s'-\ell}\,\frac{B_{m}(z)B_{\ell}(z'-z+1)}{M^{\ell+m+s+s'-2}}\;,
\end{equation}
valid when $s'\neq1$ and $s+s'\not\in2-\mathbb{N}$, and similarly when $s+s'\in2-\mathbb{N}$ from (\ref{Hurwitz zeta double tilde sum integral}) and (\ref{Hurwitz zeta double integers}).
\end{subsubsection}

\end{subsection}

\end{section}

\begin{section}{Calculation of various integrals}
\label{appendix integrals}
The various asymptotic expansions obtained in this paper using the Euler-Maclaurin formula involve integrals. Most of them have an integrand that depends on the variable of integration $u\in[0,\rho]$ only through $\Phi(u-\tfrac{\rho}{2})$, with $\Phi$ defined in (\ref{Phi}). Such integrals can be computed by making the change of variables $z=\Phi(u-\tfrac{\rho}{2})$. From (\ref{Phi'}), the Jacobian is given by
\begin{equation}
\rmd u=-\frac{\rmd z}{2\rmi\pi}\Big(\frac{\rho}{z}+\frac{1}{1-z}\Big)\;.
\end{equation}
The variable $z$ lives on the clockwise contour $\overline{\mathcal{C}_{0}}$, which starts and ends at $z=-\rho/(1-\rho)$ for $u=0$ and $u=\rho$. The contour encloses $0$ but not $1$, see figure \ref{fig contour Phi}. Hence, for the simplest integrands that do not involve branch cuts, the calculation of the integral reduces to a simple residue calculation.

In the rest of this appendix, we treat some slightly more complicated integrals on two dimensional domains, with integrands having branch cuts that cross the contour of integration. We use the notation $\mathcal{C}_{0}=\{\Phi(\tfrac{\rho}{2}-u),0\leq u\leq\rho\}$ for the counter clockwise contour corresponding to $\overline{\mathcal{C}_{0}}$.
\begin{figure}
  \begin{center}
    \includegraphics[width=100mm]{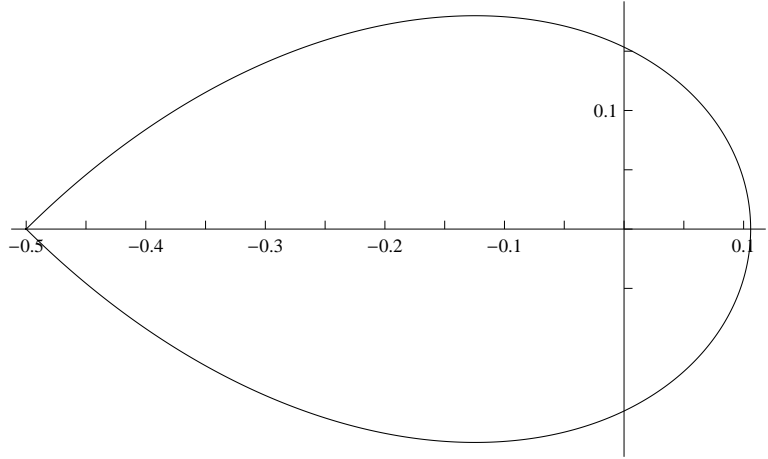}
  \end{center}
  \caption{Contour on which the Bethe roots $y_{j}$ accumulate in the complex plane for the first eigenstates in the thermodynamic limit with density of particles $\rho=1/3$. It crosses the negative real axis at $-\rho/(1-\rho)$. For any value of $\rho$, the contour encloses $0$ but not $1$.}
  \label{fig contour Phi}
\end{figure}

\begin{subsection}{A double integral}
We consider the double integral
\begin{equation}
I_{1}=\int_{0}^{\rho}\rmd u\,\int_{u}^{\rho}\rmd v\,\log\Big(-\rmi\big(\Phi(u-\tfrac{\rho}{2})-\Phi(v-\tfrac{\rho}{2})\big)\Big)\;.
\end{equation}
Making the changes of variables $z=\Phi(u-\tfrac{\rho}{2})$ and $w=\Phi(v-\tfrac{\rho}{2})$ leads to
\begin{equation}
\label{I double tmp 1}
I_{1}=\int_{\overline{\mathcal{C}_{0}}}\frac{\rmd z}{2\rmi\pi}\,\Big(\frac{\rho}{z}+\frac{1}{1-z}\Big)\int_{z}^{-\tfrac{\rho}{1-\rho}}\frac{\rmd w}{2\rmi\pi}\,\Big(\frac{\rho}{w}+\frac{1}{1-w}\Big)\log(-\rmi(z-w))\;.
\end{equation}
The inner integral can be computed in terms of the dilogarithm function $\Li_{2}$. Indeed, using $\Li_{2}'(w)=-w^{-1}\log(1-w)$, we observe that the function $F_{z}$ defined by
\begin{equation}
F_{z}(w)=\rho\Big(\Li_{2}\big(\frac{z}{w}\big)+\frac{(\log w)^{2}}{2}+\frac{\rmi\pi}{2}\,\log w\Big)
-\Big(\Li_{2}\big(\frac{1-z}{1-w}\big)+\frac{(\log(1-w))^{2}}{2}-\frac{\rmi\pi}{2}\,\log(1-w)\Big)\;
\end{equation}
verifies
\begin{equation}
F_{z}'(w)=\Big(\frac{\rho}{w}+\frac{1}{1-w}\Big)\log(-\rmi(z-w))\;.
\end{equation}
The contour of integration for $w$ in (\ref{I double tmp 1}) does not cross the branch cuts coming from the dilogarithm. The integration over $z$ can be done by taking the residue at $0$ for all the terms such that the contour does not cross a branch cut. Using $\Li_{2}(0)=0$, one finds
\begin{eqnarray}
&& I_{1}=\frac{\rho}{2}\int_{\mathcal{C}_{0}}\frac{\rmd z}{(2\rmi\pi)^{2}}\,\Big(\frac{\rho}{z}+\frac{1}{1-z}\Big)\big((\log z)^{2}+\rmi\pi\log z\big)\\
&&\hspace{4mm} +\frac{\rho}{2\rmi\pi}\Big(\frac{\rmi\pi b_{0}}{2}-\frac{(1-\rho)\pi^{2}}{6}+\frac{(1-\rho)(\log(1-\rho))^{2}}{2}-\frac{\rho(\log\rho)^{2}}{2}+\rho\log\rho\log(1-\rho)+\Li_{2}(1-\rho)\Big)\;.\nonumber
\end{eqnarray}
The last integral can be computed using the fact that the contour $\mathcal{C}_{0}$ intersects the negative real axis at $z=-\rho/(1-\rho)$ and the identities $(\log z)/z=\partial_{z}(\log z)^{2}/2$, $(\log z)^{2}/z=\partial_{z}(\log z)^{3}/3$, $(\log z)/(z-1)=\partial_{z}(\Li_{2}(z)+\log z\log(1-z))$ and $(\log z)^{2}/(z-1)=\partial_{z}(-2\Li_{3}(z)+2\log z\,\Li_{2}(z)+(\log z)^{2}\log(1-z))$. After some simplifications, one finds
\begin{equation}
I_{1}=\frac{\rho b_{0}}{2}\;,
\end{equation}
with $b_{0}$ defined in (\ref{b0}).
\end{subsection}

\begin{subsection}{Another double integral}
We consider the double integral
\begin{equation}
I_{2}=\int_{0}^{\rho}\rmd u\,\int_{u}^{\rho}\rmd v\,\log\frac{(\sqrt{u}+\sqrt{v})(\sqrt{\rho-u}+\sqrt{\rho-v})}{(v-u)(\sqrt{-\rmi u}+\sqrt{\rmi(\rho-v)})}\;.
\end{equation}
It can be rewritten as
\begin{equation}
I_{2}=\int_{0}^{\rho}\rmd u\,\int_{u}^{\rho}\rmd v\,\log(\sqrt{u}+\sqrt{v})
-\int_{0}^{\rho}\rmd u\,\int_{u}^{\rho}\rmd v\,\log(v-u)
-\int_{0}^{\rho}\rmd u\,\int_{u}^{\rho}\rmd v\,\log(\sqrt{-\rmi u}+\sqrt{\rmi(\rho-v)})\;.
\end{equation}
Using $\log(\sqrt{u}+\sqrt{v})=\partial_{v}((v-u)\log(\sqrt{u}+\sqrt{v})+\sqrt{u v}-v/2)$ and $\log(\sqrt{-\rmi u}+\sqrt{\rmi(\rho-v)})=\partial_{v}(-(u+\rho-v)\log(\sqrt{-\rmi u}+\sqrt{\rmi(\rho-v)})+\rmi\sqrt{u}\sqrt{\rho-v}-v/2)$, one finds
\begin{equation}
I_{2}=\frac{9\rho^{2}}{8}-\frac{\rho^{2}\log\rho}{4}\;.
\end{equation}
\end{subsection}
\end{section}

\end{document}